\documentstyle[aps,psfig]{revtex}

\newcommand {\ket}[2]{\ \vert \, #1 #2 \, \rangle}
\newcommand {\bra}[2]{\ \langle\, #1 #2 \, \vert}

\begin{document}
\draft
\preprint{MKPH-T-97-9}
\title{{ \bf Parity violation in quasifree electron scattering off 
the deuteron\footnote{Supported by the Deutsche Forschungsgemeinschaft 
(SFB 201)}}}

\author{G.\ K\"{u}ster\footnote{Present address: Deutsches 
Krebsforschungszentrum, Forschungsschwerpunkt 05, D-69120 Heidelberg, Germany}
 and H.\ Arenh\"ovel}
\address{
Institut f\"{u}r Kernphysik, Johannes Gutenberg-Universit\"{a}t,
       D-55099 Mainz, Germany}
\date{\today}
\maketitle

\begin{abstract}
For deuteron electroweak disintegration, parity violating effects are
investigated which arise from the interference of $\gamma$ and $Z$ 
exchange as well as from the hadronic sector via a small 
parity violating component in the deuteron. 
The general formalism for the differential cross section and polarization 
observables of electromagnetic deuteron disintegration is extended to 
incorporate parity violating contributions. Formal expressions for the 
additional structure functions are derived. Results are presented for 
these parity violating structure functions for quasifree kinematics 
neglecting final state interaction and two-body effects. 
Both types of parity violating contributions to the asymmetry of the 
inclusive reaction with respect to longitudinally polarized
electrons are evaluated. The one from 
parity violating deuteron components is negligible over the whole range 
of momentum transfers considered.  
\end{abstract}

\pacs{PACS numbers: 12.15.Ji, 13.60.-r, 24.70.+s, 24.80.+y, 25.30.Fj}

\section{Introduction} 
\label{intro}

After the discovery of parity violation in weak processes in 1957, their 
manifestation in electromagnetic and strong processes has been subject of 
many detailed investigations \cite{Hen69,FiT73,Gar73,AdH85,DuZ87,MuD94}. 
The recent interest in studying parity violation by electroweak
interference is motivated by the possibility to investigate the so-called 
strangeness or better 
$s\bar s$-content of the nucleon, a quantity of particular interest with 
respect to the nucleon spin structure functions as measured 
in deep inelastic scattering \cite{EMC,SLAC,SMC}. In fact, 
several experiments to measure parity violation in electron scattering 
off hydrogen and deuterium are presently underway 
\cite{Pro90,FiS91,Har93,BeA96}. 
In these experiments, deuterium serves 
as a neutron target and for this reason quasifree kinematics is 
preferred in order to minimize possible interaction effects. 

Parity violation in inelastic electron deuteron scattering has been studied 
before theoretically by Hwang et al.\ \cite{HwH80,HwH81} in the low energy 
and momentum transfer domain and by Hadjimichael et al.\ \cite{HPD92} for 
a larger kinematical range, in particular for quasifree kinematics at high 
momentum transfers. Relativistic contributions have been considered recently 
by Poulis \cite{Pou96a,Pou96b}. Furthermore, during completion of this work 
another calculation by Mosconi and Ricci appeared \cite{MoR97}. While in 
\cite{HwH81} the parity violation by electroweak interference as well as 
through parity violating nuclear components has been considered, the latter 
has been neglected throughout in the more recent evaluations 
\cite{HPD92,Pou96a,Pou96b,MoR97}, although the authors of \cite{HPD92} 
remark that such effects should be investigated in order to see for which 
kinematical situations they can indeed be neglected or have to be included. 
This could be of relevance for the SAMPLE experiment \cite{BeA96} which 
will measure the strange magnetic form factor $G_M^{(s)}$ at low momentum 
transfer. 

It is, therefore, the purpose of this study to investigate parity violation in
deuteron disintegration by electrons in the quasifree region for energy and 
momentum transfers far beyond the ones studied in \cite{HwH81} by considering 
simultaneously both possibilities, i.e., on one hand parity violation 
in the deuteron via a parity violating part in the $NN$ potential, and 
on the other hand in the scattering process itself through the 
electroweak interference from virtual photon and $Z$ exchange. 

The presence of a parity violating $NN$ potential $V^{pnc}$ has the 
consequence that the 
nuclear states are no longer states of good parity, i.e., besides 
the dominant state $\ \vert \,J^{\pi}\, \rangle$ of angular momentum 
$J$ and parity $\pi$ there will be a small admixture of opposite 
parity $\ \vert \,J^{-\pi}\,\rangle$, with an amplitude ${\cal F}$ of 
the order $10^{-6}$ \cite{McK69}. In the case of the deuteron, the 
admixture of opposite parity components must result in a $P$-wave, 
given the possible spin states and the total angular momentum. 
For the calculation of these $P$-wave components we will use for 
$V^{pnc}$ the 
one-boson-exchange model of Desplanques, Donoghue and Holstein \cite{DDH80}. 
Based on the quark model and $SU(6)_w$ symmetry, they made 
predictions for all meson-nucleon couplings from both charged and neutral 
current pieces of the weak Hamiltonian. 

This paper is organized as follows: In Sect.\ \ref{formew} we
extend the basic formalism for describing pure electromagnetic 
electrodisintegration of the deuteron to the electroweak case, and introduce 
additional parity violating structure functions. In Sect.\ \ref{hadrpv} 
the parity violating $NN$ potential is introduced and the parity 
violating components to the deuteron wave function are calculated. 
Furthermore, we show a few of the resulting new pnc (parity nonconserved) 
structure functions for the quasifree case. The pnc structure functions 
corresponding to the second mechanism of parity violation, i.e., electroweak
interference, are considered in Sect.\ \ref{ewint}. The quantity of
experimental interest, the longitudinal asymmetry of the inclusive reaction, 
is investigated for both cases 
in Sect.\ \ref{asym}. Finally, we summarize and give some conclusions 
in Sect.\ \ref{concl}. 


\section{Formalism for electroweak disintegration}
\label{formew}

In this section we briefly present the basic formalism for the
electroweak disintegration of the deuteron. We follow closely the formalism and 
notation for electrodisintegration including polarization effects as outlined 
in Ref.~\cite{ALT93} and extend it to the electroweak case.

The general expression for the cross section including beam and target 
polarization is given by 
\begin{eqnarray}
\label{GL1}
d\sigma_{fi} & = & (2\pi)^{-5} 
\delta^{(4)}(p_{n}+p_{p}-q-P_d)
\,tr({\cal M}^\dagger_{fi}{\cal M}_{fi}\hat \rho^e \hat \rho^d) 
\,\frac{m_e^2\,d^{3}k_2}{4k_{1\, 0}k_{2\,0}} 
\,\frac{M^{2}}{M_{d}}
\,\frac{d^3p_n}{E_{n}}\frac{d^3p_p}{E_{p}}\,.
\end{eqnarray}
The momenta of the initial and the scattered electrons 
are denoted by $k_1$ and $k_2$, respectively, while
$q_{\mu}^2=q_0^2-\vec q^{\, 2}$ is the four momentum transfer squared 
($q=k_1-k_2$). The deuteron and final proton and neutron momenta are 
denoted by $P_d$ and $p_{p/n}$ and their masses by $M$ and $M_d$, 
respectively. The density matrices $\hat \rho^e$ and $\hat \rho^d$ 
describe possible beam and target polarization. 
Covariant normalization has been assumed, i.e., $(2\pi)^{3} E/m$
for fermions and $(2\pi)^{3}2E$ for bosons. 

\begin{figure}
\centerline{\psfig{figure=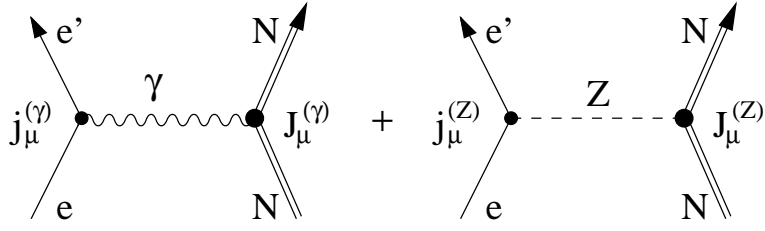,width=10cm,angle=0}}
\vspace*{.5cm}
\caption{Diagrams for $\gamma$ and $Z$ exchange in electroweak electron 
hadron scattering.
}
\label{fig1}
\end{figure}

The amplitude ${\cal M}_{fi}$ 
contains in the lowest order contributions from both virtual $\gamma$ 
and $Z$ exchange (see Fig.\ \ref{fig1}) with the latter naturally being 
strongly suppressed since we restrict ourselves to the low 
momentum transfer region ($-q_\mu^2 \ll M^2_Z$).  
The invariant matrix element in the
one-boson-exchange approximation thus contains two contributions 
\cite{MuD92}
\begin{eqnarray}
 {\cal M}_{fi}&=& \frac{e^2}{q_\mu^2}\ j^{(\gamma)\,\mu}J_{fi,\,\mu}^{(\gamma)}
 + \sqrt{2}G_F \,j^{(Z)\,\mu}J_{fi,\,\mu}^{(Z)} \,.
\end{eqnarray}
Here and in the following, the superscripts $\gamma$ and $Z$ indicate the 
electromagnetic and weak neutral current contributions. The lepton and 
hadron currents are denoted by $j^{(\gamma/Z)}_{\mu}$ and 
$J_{fi,\,\mu}^{(\gamma/Z)}$, respectively. Furthermore, $e$ denotes the 
elementary charge with $\alpha=e^2/4\pi$ as fine structure constant, and 
the weak Fermi coupling constant $G_F$ is given by 
\begin{eqnarray} 
\frac{G_F}{\sqrt{2}} & = &  \frac{g^2}{8\cos^2\theta_W M_Z^2} \,,
\end{eqnarray}
where $g$ denotes the electroweak coupling constant, $\theta_W$ the Weinberg 
angle, and $e=g\sin \theta_W$. In the following we use $\sin^2 \theta_W=0.232$. 
The lepton currents are defined by 
\begin{eqnarray}
j^{(\gamma)\,\mu}&=& \bar u(k_2)\gamma^\mu u(k_1)\,,\\
j^{(Z)\,\mu}&=& \bar u(k_2)\gamma^\mu (g_v + g_a \gamma_5)u(k_1)\,,
\end{eqnarray}
with
\begin{eqnarray}
g_{v}&=&-\frac{1}{2} + 2\sin^2\theta_W\,,\\
g_{a}&=& \frac{1}{2}\,.
\end{eqnarray}
The hadronic currents are given later in Sect.\ \ref{ewint}. Note, that our 
expressions for the neutral currents contain an additional factor $1/2$ 
compared to Ref.\ \cite{MuD92}.

Allowing for longitudinal electron polarization of degree $h$, one finds 
the well known expression \cite{MuD94}
\begin{eqnarray}
m_e^2\,tr({\cal M}^\dagger_{fi}{\cal M}_{fi}\hat \rho^e \hat \rho^d)
&=& \Big(\frac{e^2}{q_\mu^2}\Big)^2 
\eta_{\mu\nu}^{\gamma \gamma}(h)\,W^{\gamma \gamma,\,\mu\nu}_{fi}(\hat \rho^d)
+ 2 G_F^2\,\eta_{\mu\nu}^{Z Z}(h)\,
W^{Z Z,\,\mu\nu}_{fi}(\hat \rho^d)\nonumber\\
&&+\sqrt{2} G_F\, \frac{e^2}{q_\mu^2} \,\eta_{\mu\nu}^{\gamma Z}(h) \,
(W^{\gamma Z,\,\mu\nu}_{fi}(\hat \rho^d) 
+ W^{Z \gamma,\,\mu\nu}_{fi}(\hat \rho^d))\,,
\label{traceM}
\end{eqnarray}
where the various lepton tensors are given by
\begin{eqnarray}
\eta_{\mu\nu}^{\gamma \gamma}(h) &=& \eta_{\mu\nu}^{v v}(h)\,,\\
\eta_{\mu\nu}^{\gamma Z}(h) &=& 
g_v \eta_{\mu\nu}^{v v}(h) + g_a \eta_{\mu\nu}^{v a}(h)\,,\\
\eta_{\mu\nu}^{Z Z}(h) &=& (g_v^2 + g_a^2) \eta_{\mu\nu}^{v v}(h) 
 + 2 g_v g_a \eta_{\mu\nu}^{v a}(h) \,.
\end{eqnarray}
Since the $Z$ couples to both the vector and the
axial vector current (with couplings $g_v$ and $g_a$), one has, compared to 
the pure electromagnetic case, now two types 
of lepton tensors $\eta_{\mu\nu}^{v v}$ and 
$\eta_{\mu\nu}^{v a}$, where the latter arises from the 
interference of the vector with the axial vector current, 
\begin{eqnarray}
\eta_{\mu\nu}^{v v}(h) &=& \eta_{\mu\nu}^0 + h \eta_{\mu\nu}^{\prime}\,,\\
\eta_{\mu\nu}^{v a}(h) &=& \eta_{\mu\nu}^{\prime} + h \eta_{\mu\nu}^{0}\,.
\end{eqnarray}
In the high energy limit, i.e., electron mass $m_e=0$, one has 
\begin{eqnarray}
\eta_{\mu\nu}^0 &=& (k_{1\,\mu} k_{2\,\nu} + k_{2\,\mu} k_{1\,\nu})
- g_{\mu \nu} k_1\cdot k_2\nonumber\\
&=& \frac{1}{2}(k_\mu k_\nu - q_\mu q_\nu + g_{\mu\nu} q_\rho^2)
\,,\\
\eta_{\mu\nu}^{\prime} &=& i \varepsilon_{\mu \nu \alpha \beta} 
k_1^{\alpha} k_2^{\beta}\nonumber\\
&=& \frac{i}{2} \varepsilon_{\mu \nu \alpha \beta}k^\alpha q^\beta\,,
\end{eqnarray}
where $k=k_1+k_2$. 
The hadronic tensors appearing in (\ref{traceM}) are given by the 
electromagnetic and weak current matrix elements, for example, 
\begin{eqnarray}
W^{\gamma \gamma,\,\mu\nu}_{fi}(\hat \rho^d) 
&=& tr(J_{fi}^{(\gamma)\, \mu\,\dagger} J_{fi}^{(\gamma)\, \nu}\hat \rho^d)\,,
\end{eqnarray}
where the trace refers to the deuteron spin quantum numbers. 

Proceeding as in the electromagnetic
case, one obtains in analogy to
the pure electromagnetic process the following expression for the 
differential cross section including both beam and target polarization
(for details see \cite{Kue96}) 
\begin{eqnarray}
\frac{d^{3}\sigma^{\gamma +Z}}{dk_2^{lab} d \Omega_{k_2}^{lab} 
d\Omega_{np}^{c.m.}} &=&  \frac{\alpha}{2 \pi^2} 
\frac{ k_2^{lab}}{k_1^{lab}q_\mu^4} 
\sum_{\lambda \lambda ^{\prime} s m_s m_d m_d^{\prime}} 
\rho_{m_d m_d^{\prime}}^d  \Big( \rho_{\lambda \lambda ^{\prime}}^{vv}  
T_{s m_s \lambda m_d}^{\gamma} T_{s m_s \lambda ^{\prime} 
m_d^{\prime}}^{{\gamma}\ \ast} 
\nonumber \\
 & & \hspace{4em}{}+ \frac{G_F}{\sqrt{2}}\frac{q_\mu^2}{\pi \alpha} 
  (g_v \rho_{\lambda \lambda ^{\prime}}^{vv} 
  +g_a \rho_{\lambda \lambda ^{\prime}}^{va})
  \,\Re e\, \Big[  T_{s m_s \lambda m_d}^{\gamma}
  T_{s m_s \lambda ^{\prime} m_d^{\prime}}^{{Z}\ \ast} \Big]
  \Big)\ ,
\label{sig_interf}
\end{eqnarray}
where the terms quadratic in the weak amplitude 
($T^{Z\,\ast}_\mu T^{Z}_\nu$) have been omitted since
they are of order ${{\cal O}} (q_\mu^4/M_Z^4)$ compared to the
electromagnetic process. 
The direction of the outgoing proton in the final $np$ 
c.m.\ system is denoted by $\Omega_{np}$. One finds for the
spherical components of the two types of virtual boson density matrices 
\begin{eqnarray}
 \rho_{\lambda \lambda^{\prime}}^{vv} &=&  
\rho_{\lambda \lambda^{\prime}}^0 + h 
 \rho_{\lambda \lambda^{\prime}}^{\prime} \ ,\\
 \rho_{\lambda \lambda^{\prime}}^{va} &=&  
\rho_{\lambda \lambda^{\prime}}^{\prime} + h 
 \rho_{\lambda \lambda^{\prime}}^0 \ .
\end{eqnarray}
They obey the symmetry relations
\begin{eqnarray}
 \rho_{\lambda \lambda^{\prime}}^{vv/va} & = & 
\rho_{\lambda^{\prime} \lambda}^{vv/va} \ ,\\
 \rho_{-\lambda -\lambda^{\prime}}^0 & = & (-)^{\lambda +\lambda^{\prime}}
    \rho_{\lambda \lambda^{\prime}}^0 \ , \\
 \rho_{-\lambda -\lambda^{\prime}}^{\prime} & = & 
 (-)^{\lambda +\lambda^{\prime}+1}\rho_{\lambda \lambda^{\prime}}^{\prime} \ .
\end{eqnarray}
The nonvanishing components are 
\begin{eqnarray}
 \rho_L=\rho_{00}^0=-\beta^2 q_{\nu}^2\frac{\xi^2}{2\eta} 
\,&,&\quad \rho_T=\rho_{11}^0
  =-\frac{1}{2}q_{\nu}^2\,\Big(1+\frac{\xi}{2 \eta} \Big) \ ,\\
 \rho_{LT}=\rho_{01}^0=-\beta q_{\nu}^2 \frac{\xi}{\eta}\,
 \sqrt{\frac{\eta+ \xi}{8}}
\, &,&\quad \rho_{TT}=\rho_{-11}^0=q_{\nu}^2\frac{\xi}{4 \eta} \\
 \rho_{LT}^{\prime}=\rho_{01}^{\prime}=
 -\frac{1}{2}\,\beta\frac{q_{\nu}^2}{\sqrt{2\eta}}\,\xi \,&,& \quad
 \rho_T^{\prime}=\rho_{11}^{\prime}=
  -\frac{1}{2}q_{\nu}^2\, \sqrt{\frac{\eta+\xi}{\eta}} \ ,
\end{eqnarray}
with 
\begin{eqnarray}
\beta = {|{\vec q}^{\,lab}| \over |{\vec q}^{\,c}|}\,,\quad
\xi = -\frac{q_{\nu}^2}{|{\vec q}^{\,lab}|^2} \,, \quad 
\eta = \tan^2\frac{\theta_e}{2}\ ,
\end{eqnarray}
where $\beta$ expresses the boost from the lab system to the frame in which 
the hadronic tensor is evaluated and ${\vec q}^{\,c}$ denotes the 
momentum transfer in this frame. In this work, we take the c.m.\ frame of 
the final $np$ state for the evaluation, and the momentum transfer in this 
frame will be denoted by $\vec q$ throughout. In order to make contact to 
the kinematic functions $v_{\alpha^{(\prime)}}$ in the review of Musolf et 
al.\ \cite{MuD94}, we note the simple relation
\begin{eqnarray}
\rho_\alpha^{(\prime)} &=& -\frac{q_\mu^2}{2\eta}\, v_{\alpha^{(\prime)}}\,,
\end{eqnarray}
where $\alpha = L,\, T,\, LT,\, TT$.

In Eq.\ (\ref{sig_interf}), the $T$-matrices were introduced, which are 
related to the electromagnetic and neutral current matrix elements between 
intrinsic states by
\begin{eqnarray}
 T_{s m_s \lambda m_d}^{\gamma} & = & -\pi \sqrt{2 \alpha p_{np}E E_d/M_d}
 \ \langle\,s m_s \, \vert J_{\lambda}^{\gamma}(\vec{q},\vec{Q}) \ 
\vert \,m_d\, \rangle \ ,\label{tgamma}\\
 T_{s m_s \lambda m_d}^{Z} & = & -\pi \sqrt{2 \alpha p_{np}E E_d/M_d}
 \bra{s}{m_s} J_{\lambda}^{Z}(\vec{q},\vec{Q}) \ket{}{m_d} \ ,\label{tzet}
\end{eqnarray}
where $\lambda = \pm$ refer to the transverse current components (with 
respect to $\vec q\,$), while the $\lambda = 0$ component is given by
\begin{eqnarray}
J_0 &=& -\frac{|\vec q\,|^2}{q_\mu^2}(\rho - \frac{\omega}{|\vec q\,|^2} 
\vec q \cdot \vec J)\nonumber\\
&=& \rho - \frac{\omega }{q_\mu^2}(\omega\rho - \vec q \cdot \vec J)\,,
\end{eqnarray}
which reduces to the charge density $\rho$ for a conserved current. 
Furthermore, $E_d$ and $E$ denote the deuteron and nucleon energies 
in the $np$ c.m.\ frame, respectively. Here, the c.m.\ motion of the 
initial and final hadronic states with c.m.\ momenta $\vec P_i$ and 
$\vec P_f$, respectively, has been eliminated and we have switched to 
noncovariant normalization. One should note 
that the intrinsic current operators depend also on 
$\vec Q=\vec P_i +\vec P_f$. 

The final intrinsic state is characterized by the relative $np$-momentum
$\vec{p}_{np}$, the spin $s$ and its projection $m_s$ for which 
the direction of $\vec{p}_{np}$ is chosen as quantization
axis, whereas for the initial deuteron state the momentum $\vec{q}$ of
the virtual boson is chosen as quantization axis. The direction of
$\vec{p}_{np}$ is described by the angles $\theta$ and $\phi$ with
respect to a coordinate system whose $z$-axis is chosen parallel to 
$\vec{q}$ and the $y$-axis parallel to $\vec k_1 \times \vec k_2$ 
(for details see Fig.\ \ref{kine} and \cite{ALT93}). 

Then the $\phi$-dependence of the $T$-matrices can be separated so that the 
reduced $t$-matrices depend on $\theta$ and $|\vec q\,|$ only 
\begin{eqnarray}
 T_{s m_s \lambda m_d}^{{\gamma/Z}} (\theta,\phi) & = & e^{i(\lambda+m_d)\phi}
                \ t_{s m_s \lambda m_d}^{{\gamma/Z}}(\theta)\ .
\end{eqnarray}
The electromagnetic current matrix element can be split into a leading, 
parity conserving part ($\gamma_v$) and a very small contribution 
($\gamma_a$, analogue to an axial current) from parity 
violating components of the wave functions 
\begin{equation}
t_{s m_s \lambda m_d}^{\gamma}(\theta)=t_{s m_s \lambda m_d}^{\gamma_v}(\theta)
+t_{s m_s \lambda m_d}^{\gamma_a}(\theta)\,.
\end{equation}
For the parity conserved part, the transformation properties under space and
time inversion lead to the following symmetry property for the reduced
$t$-matrices
\begin{equation}
t_{s -m_s -\lambda -m_d}^{\gamma_v}  =  (-)^{1+s+m_s+\lambda+m_d}\ 
 t_{s m_s \lambda m_d}^{\gamma_v} \ ,
\label{t_symm}
\end{equation}
while for the parity violating part one finds
\begin{equation}
t_{s -m_s -\lambda -m_d}^{\gamma_a}  =  (-)^{s+m_s+\lambda+m_d}\
 t_{s m_s \lambda m_d}^{\gamma_a} \ .
\label{t_sympviol}
\end{equation}
Just as for the leptonic current, one has two contributions for the weak
hadronic current, namely a vector and an axial vector piece
\begin{equation}
t_{s m_s \lambda m_d}^{Z}(\theta)=t_{s m_s \lambda m_d}^{Z_v}(\theta)
+t_{s m_s \lambda m_d}^{Z_a}(\theta)\,.
\end{equation}
The reduced $t$-matrices $t_{s m_s \lambda m_d}^{Z_{v/a}}$ have the same 
symmetry properties as in (\ref{t_symm}) for $v$ and in (\ref{t_sympviol}) 
for $a$. 

Now one can proceed exactly as in \cite{ALT93} by 
inserting the deuteron density matrix, which is characterized by
vector and tensor polarization parameters $P_1^d$ and $P_2^d$, 
respectively, and by the angles $\theta_d$ and $\phi_d$, describing the
direction of the orientation axis $\hat d$ of the polarized deuteron 
target with respect to the coordinate system associated with the 
three-momentum transfer $\vec q$ (see \cite{ALT93} for details), 
and exploiting the symmetry 
properties of the boson density matrices and the reduced $t$-matrices. 
It is useful, in order to distinguish the contributions according to their 
parity properties, to introduce the following two index sets 
\begin{eqnarray}
{\cal S}_v&:=&\{\gamma_v,\,Z_v\}\,,\\
{\cal S}_a&:=&\{\gamma_a,\,Z_a\}\,,
\end{eqnarray}
because, as will be apparent below, the contributions to the cross section 
from the same set have formally the same type of structure functions. 

In Ref.\ \cite{ALT93}, all observables had been classified according to 
their parity properties, i.e., a set $A$ for the scalar and a set $B$ for 
the pseudoscalar observables. To each set belongs a corresponding set of 
structure functions. Having now two types of $t$-matrices ($v$- and 
$a$-type) it is obvious that the combinations $v$-$v$ and $a$-$a$ lead to 
the same type of structure functions as in \cite{ALT93}, whereas for the 
$v$-$a$ combinations the role of the $A$- and $B$-type observables is 
interchanged. Keeping this in mind, one finds the following compact 
expression for the differential cross section 
\begin{eqnarray}
 S(h,P_1^d,P_2^d) &=& 
\frac{d^{3}\sigma^{\gamma +Z}}{dk_2^{ lab} d \Omega_{k_2}^{ lab} 
d\Omega_{np}^{ c.m.}} \nonumber\\
&=&
 \bar S(\gamma_v;P_1^d,P_2^d)+\bar S(\gamma_a;P_1^d,P_2^d)\nonumber\\
&&
 +a_v\Big(\bar S(Z_v;P_1^d,P_2^d)+\bar S(Z_a;P_1^d,P_2^d)\Big)\nonumber\\
&&
 +a_a\Big(\bar S'(Z_v;P_1^d,P_2^d)+\bar S'(Z_a;P_1^d,P_2^d)\Big)
\nonumber\\
&&
+h\Big[
 \bar S'(\gamma_v;P_1^d,P_2^d)+\bar S'(\gamma_a;P_1^d,P_2^d)\nonumber\\
&&
 +a_v\Big(\bar S'(Z_v;P_1^d,P_2^d)+\bar S'(Z_a;P_1^d,P_2^d)\Big)\nonumber\\
&&
 +a_a\Big(\bar S(Z_v;P_1^d,P_2^d)+\bar S(Z_a;P_1^d,P_2^d)\Big)\Big]\,,
\label{s-gam-2}
\end{eqnarray}
where we have defined
\begin{equation}
a_{v/a}=\frac{G_F}{\sqrt{2}}\frac{q_\mu^2}{\pi \alpha} g_{v/a}\,.
\end{equation} 
The abnormal parity admixture in the 
deuteron has been considered for photon exchange only. 
Here, we have introduced for $C\in {\cal S}_v\cup {\cal S}_a$, where $C$ 
characterizes the different current contributions, 
\begin{eqnarray}
\bar S(C;\,P_1^d,P_2^d)
&=&
 c \sum _{I=0}^2 P_I^d \sum _{M=0}^I
 \Bigl\{  (\rho _L f_L^{IM,C} + \rho_T f_T^{IM,C} +
 \rho_{LT} {f}_{LT}^{IM+,C} \cos \phi \nonumber\\
& &
 + \rho _{TT} {f}_{TT}^{IM+,C} \cos2 \phi)
 \cos (M\tilde{\phi}-\bar\delta_{I}^{C} {\pi \over 2})\nonumber \\
& & 
-(\rho_{LT} {f}_{LT}^{IM-,C} \sin \phi
+ \rho _{TT} {f}_{TT}^{IM-,C} \sin2 \phi)
\sin (M\tilde{\phi}-\bar\delta_{I}^{C} {\pi \over 2})
\Big\} d_{M0}^I(\theta_d)\,,\label{obsfin}\\
\bar S'(C;\,P_1^d,P_2^d)
&=&
 c \sum _{I=0}^2 P_I^d \sum _{M=0}^I
 \Bigl\{ (\rho'_T f_T^{\prime IM,C}
 + \rho '_{LT} {f}_{LT}^{\prime IM-,C} \cos \phi ) 
 \sin (M\tilde{\phi}-\bar\delta_{I}^{C} {\pi \over 2}) \nonumber\\
& &
 \qquad + \rho '_{LT} {f}_{LT}^{\prime IM+,C} \sin \phi
 \cos (M\tilde{\phi}-\bar\delta_{I}^{C} {\pi \over 2})
 \Big\} d_{M0}^I(\theta_d)\,,\label{obsfins}
\end{eqnarray}
with
\begin{equation}
c = {\alpha \over 6 \pi^2} {k^{lab}_2 \over k_1^{lab} q_{\nu}^4}
= -\frac{\eta}{2q_{\nu}^2}\frac{\sigma_M}{3\pi^2 \alpha}\,,
\end{equation}
where $\sigma_M$ denotes the Mott cross section, 
$\tilde \phi=\phi -\phi_d$ and
\begin{equation}
\bar\delta_{I}^{C}:=\left\{\matrix{\delta_{I1} & \mbox{for}\; 
C\in {\cal S}_v \cr 1-\delta_{I1} & 
\mbox{for}\; C\in {\cal S}_a \cr} \right\}\,.\label{deltaic}
\end{equation}
For $C\in{\cal S}_v$, the structure functions 
correspond to those of the $A$-type observables in \cite{ALT93} while for 
$C\in{\cal S}_a$, they correspond to the ones of 
the $B$-type observables as already mentioned above. 

The various structure functions which appear in (\ref{obsfin}) and 
(\ref{obsfins}) are defined by 
\begin{eqnarray}
f_{L}^{IM,C}&=&\frac{2}{1+\delta_{M0}}\Re e\Big[i^{\bar \delta^C_I}
v_{00 I M}^{C}\Big]\,,
\label{ffirst}\\
f_{T}^{IM,C}&=&\frac{4}{1+\delta_{M0}}\Re e\Big[i^{\bar \delta^C_I}
v_{11 I M}^{C}\Big]\,,\\
f_{LT}^{IM\pm,C}&=&\frac{4}{1+\delta_{M0}}\Re e\Big[i^{\bar \delta^C_I}
(v_{01 I M}^{C}\pm(-)^{I+M+\delta_{C}}
v_{01 I -M}^{C})\Big]\,,\\
f_{TT}^{IM\pm,C}&=&\frac{2}{1+\delta_{M0}}\Re e\Big[i^{\bar \delta^C_I}
(v_{-11 I M}^{C}\pm(-)^{I+M+\delta_{C}}
v_{-11 I -M}^{C})\Big]\,,\\
f_{T}^{\prime IM,C}&=&\frac{4}{1+\delta_{M0}}\Re e\Big[i^{1+\bar \delta^C_I}
v_{11 I M}^{C}\Big]\,,\\
f_{LT}^{\prime IM\pm,C}&=&\frac{4}{1+\delta_{M0}}\Re e\Big[i^{1+\bar \delta^C_I}
(v_{01 I M}^{C}\pm(-)^{I+M+\delta_{C}}
v_{01 I -M}^{C})\Big]\,,
\label{flast}
\end{eqnarray}
where $M\ge 0$ and $C\in {\cal S}_v\cup {\cal S}_a$. Furthermore, we have 
introduced
\begin{equation}
\delta_{C}:=\left\{\matrix{0 & \mbox{for}\;
C\in {\cal S}_v \cr 1 &
\mbox{for}\; C\in {\cal S}_a \cr} \right\}\,.
\end{equation}
For completeness we list for all polarization observables 
the corresponding structure functions in the Appendix. 

Note that some of the structure functions vanish identically. First of all, 
for those with $M=0$ one finds from the above definitions 
\begin{eqnarray}
f_{\alpha}^{I0-,C} &=& 0\,\,\, \mbox{for}\,\,\, 
(-)^{I+\delta_{C}}=1\,,\label{zero1}\\
f_{\alpha}^{I0+,C} &=& 0 \,\,\,\mbox{for}\,\,\, 
(-)^{I+\delta_{C}}=-1\,,\label{zero2}
\end{eqnarray}
for $\alpha=L,\,T,\,LT,\,TT$. Additional vanishing structure functions 
follow from the symmetry properties of $v_{\lambda \mu IM}^C$ discussed below.

The quantities $v_{\lambda \mu IM}^C$ are given in terms of the $t$-matrix 
elements by 
\begin{eqnarray}
v_{\lambda \mu IM}^C(\theta)=\frac{\hat I \sqrt 3}{1+\delta_{C, \gamma_v}} 
&& \sum_{m m'} (-)^{1 - m} 
\left( \matrix {1&1&I \cr m & -m' &-M \cr} \right) \nonumber\\
&& \sum_{sm_s} \Big[(t^{\gamma_v}_{sm_s \lambda m'}(\theta) )^\ast
t_{sm_s \mu m}^C(\theta)+(t^{C}_{sm_s \lambda m'}(\theta))^\ast
t_{sm_s \mu m}^{\gamma_v}(\theta)\Big]\;.\label{vlam}
\end{eqnarray}
They obey two symmetry relations. The first one follows from (\ref{t_symm}) 
and (\ref{t_sympviol}) 
\begin{eqnarray}
v^C_{-\lambda -\mu I-M}&=&(-)^{I-M+\lambda -\mu +\delta_C}\;v^C
_{\lambda \mu IM} \,,\label{s1}
\end{eqnarray}
and the second one from complex conjugation
\begin{eqnarray}
(v^C_{\lambda \mu IM})^*&=&(-)^{M}v^C_{\mu \lambda I-M}\,.\label{comcon}
\end{eqnarray}
These relations may be combined to give
\begin{eqnarray} 
(v^C_{\lambda \mu IM})^*&=&(-)^{I+\lambda -\mu +\delta_C}v^C_{-\mu -\lambda IM} 
\,.\label{s2}
\end{eqnarray}
In particular, one finds from (\ref{comcon})
\begin{eqnarray}
(v^C_{\lambda \lambda IM})^*&=&(-)^{M}v^C_{\lambda \lambda I-M}\,,\label{sel1}
\end{eqnarray}
implying that $v^C_{\lambda \lambda I0}$ is real, and from (\ref{s2})
\begin{eqnarray}
(v^C_{\lambda -\lambda IM})^*&=&(-)^{I+\delta_C}v^C_{\lambda -\lambda IM}
\,,\label{sel2}
\end{eqnarray}
from which follows that $v^C_{\lambda -\lambda IM}$ is real or imaginary 
for $(-)^{I+\delta_C}=1$ or $-1$, respectively. 

The structure functions without target polarization ($(IM)=(00)$) are then 
for $C\in {\cal S}_v$ with the notation 
$f_\alpha^{(\prime) C}\equiv f_\alpha^{(\prime) 00+,C}$, and using
(\ref{sel1})
\begin{eqnarray}
 f_L^{C}=v_{0000}^{C} \,&,& \qquad f_T^{C} = 2v_{1100}^{C} \ ,\label{fcv}\\
 f_{LT}^{C}=4 \Re e(v_{0100}^{C})  \,&,& 
 \qquad f_{TT}^{C}=2v_{-1100}^{C} \ ,\\
 f_T^{\prime\, C} = 0 \,&,& \qquad 
 f_{LT}^{\prime\,C} = -4 \Im m (v_{0100}^{C}) \,.
\end{eqnarray}
Analogously, one finds for the parity violating structure functions 
($C\in {\cal S}_a$) without target polarization with the notation
$f_\alpha^{(\prime) C}\equiv f_\alpha^{(\prime) 00-,C}$, and using 
again (\ref{sel1})
\begin{eqnarray}
 f_L^{C}=0  \, &,& \qquad f_T^{C} = 0\, ,\\
 f_{LT}^{C}=-4 \Im m (v_{0100}^{C}) \,&,& \qquad 
 f_{TT}^{C}=-2 \Im m (v_{-1100}^{C})\, ,\\
 f_T^{\prime\,C} = -2 v_{1100}^{C} \,&,& 
\qquad f_{LT}^{\prime\,C}=-4 \Re e(v_{0100}^{C})\,.\label{fca}
\end{eqnarray}
Then we find for the differential coincidence cross section without target 
polarization
\begin{eqnarray}
S(h,0,0) =
 c \Big\{ &&\rho _L (f_L^{\gamma_v}+ a_v f_L^{Z_v}) 
 + \rho_T ( f_T^{\gamma_v} + a_v f_T^{Z_v})
\nonumber\\
 + &&\rho_{LT} \Big( (f_{LT}^{\gamma_v} + a_v f_{LT}^{Z_v}) \cos \phi 
                 + (f_{LT}^{\gamma_a} + a_v f_{LT}^{Z_a}) \sin \phi\Big)
\nonumber\\
 + &&\rho_{TT} \Big( (f_{TT}^{\gamma_v} + a_v f_{TT}^{Z_v}) \cos 2 \phi
                 + (f_{TT}^{\gamma_a} + a_v f_{TT}^{Z_a}) \sin 2 \phi\Big)
\nonumber\\
 - &&\rho'_T a_a f_T^{\prime Z_a} 
 - \rho'_{LT} a_a \Big(f_{LT}^{\prime Z_a}\cos \phi 
                  -f_{LT}^{\prime Z_v}\sin \phi\Big)
\nonumber\\
 +h \Big[&&\rho _L a_a f_L^{Z_v} + \rho_T a_a f_T^{Z_v}
 + \rho_{LT} a_a (f_{LT}^{Z_v} \cos \phi 
                 + f_{LT}^{Z_a} \sin \phi )
\nonumber\\
 + &&\rho_{TT} a_a (f_{TT}^{Z_v} \cos 2 \phi
                 + f_{TT}^{Z_a} \sin 2 \phi )
- \rho'_T (f_T^{\prime \gamma_a} + a_v f_T^{\prime Z_a})
\nonumber\\
- &&\rho'_{LT} \Big((f_{LT}^{\prime \gamma_a} 
                     +a_v f_{LT}^{\prime Z_a})\cos \phi
       -(f_{LT}^{\prime \gamma_v} + a_v f_{LT}^{\prime Z_v})\sin \phi
\Big)\Big]
\Big\}\,.\label{diffcross}
\end{eqnarray}
It is apparent that parity violation in the hadronic sector ($\gamma_a$) 
and via the weak axial hadron current ($Z_a$) lead to the same type of 
structure functions so that they always appear together in the combination 
$f_{\alpha}^{(\prime) \gamma_a}+a_v f_{\alpha}^{(\prime) Z_a}$ 
(see \cite{MuD94}). Similarly, 
the vector contributions appear always in the form 
$f_{\alpha}^{(\prime) \gamma_v}+a_v f_{\alpha}^{(\prime) Z_v}$. We note in 
passing, that such an expression without the $\gamma_a$-contributions has 
also been derived in \cite{MoR97}. 

The inclusive cross section for an unpolarized target, denoted by 
$\Sigma(h)$, is then obtained by integration over $\Omega_{np}$ of the 
outgoing nucleon
\begin{eqnarray}
\Sigma(h)&=& d^{2}\sigma^{\gamma +Z}/dk_2^{lab} d \Omega_{k_2}^{lab}
= \int d\Omega_{np} \ S(h,0,0)\nonumber\\
&=& c \Big\{ \rho _L (F_L^{\gamma_v}+ a_v F_L^{Z_v}) 
 + \rho_T ( F_T^{\gamma_v} + a_v F_T^{Z_v})
-\rho'_T a_a F_T^{\prime Z_a} 
\nonumber\\
&& +h \Big[\rho _L a_a F_L^{Z_v} + \rho_T a_a F_T^{Z_v}
- \rho'_T (F_T^{\prime \gamma_a} + a_v F_T^{\prime Z_a}) \Big]
\Big\}\,,\label{inclusive}\\
&=& \frac{\sigma_M}{3\pi^2\alpha}\Big\{ v_L (F_L^{\gamma_v}+ a_v F_L^{Z_v})
 + v_T ( F_T^{\gamma_v} + a_v F_T^{Z_v})
-v_{T'} a_a F_T^{\prime Z_a}
\nonumber\\
&& +h \Big[v_L a_a F_L^{Z_v} + v_T a_a F_T^{Z_v}
- v_{T'} (F_T^{\prime \gamma_a} + a_v F_T^{\prime Z_a}) \Big]
\Big\}\,,\label{inclusivea}
\end{eqnarray}
where in the second form (\ref{inclusivea}) we have switched to the 
notation of \cite{MuD94}. The inclusive form factors are defined by 
\begin{eqnarray}
 F_{L/T}^{(\prime) \,C} & = & \int d\Omega_{np}
\,f_{L/T}^{(\prime) \,C}\, .
\end{eqnarray}
The expression for the inclusive cross section reflects the well known fact 
\cite{MuD94} that the helicity dependent part is a direct measure of 
the total parity violating contributions from hadronic parity violation and 
from electroweak interference.


\section{Hadronic parity violation} 
\label{hadrpv}

A weak parity violating $NN$ interaction $V^{pnc}$ is in principle
always present in addition to the strong and parity conserving $NN$ 
interaction $V^{pc}$, with an estimated relative strength of the order 
$V^{pnc}/V^{pc}\approx 10^{-7} $. Since the strong $NN$ interaction is highly
repulsive at small distances (hard core), the effect of a direct
zero range weak $NN$ interaction is negligible. In analogy to the 
strong $NN$ interaction, one therefore assumes a one-boson-exchange 
model for the weak $NN$ interaction. In this 
case, one of the meson-nucleon vertices is parity violating
and associated with the weak interaction, while the other (strong)
vertex is parity conserving. With this assumption, the whole physics is
contained in the weak coupling constants.

Due to the presence of the hard core, one can neglect the
exchange of mesons with mass greater than the $\rho$-mass $m_{\rho}$, 
since they lead to a very short range potential. As shown by Barton
\cite{Bar61}, the parity violating potential cannot be reduced to
one-pion-exchange (OPE) alone, if only strangeness conserving weak currents are
included and $CP$ invariance is assumed. This argumentation can be
extended to the exchange of any pseudoscalar meson, forbidding the
exchange of any neutral pseudoscalar meson (e.g.\ $\pi^0$) and yielding
an isovector component $(\Delta I=1)$ in the Hamiltonian for the
charged pseudoscalar mesons \cite{Hen69}. Assuming a specific model
for the weak interaction, for example, in the GWS-model, one can derive 
from the transformation properties under isospin rotations that the 
neutral currents dominate the isovector component of the Hamiltonian 
\cite{AdH85}. Thus -- and due to the fact that in some experiments
$\Delta I=1$ is completely absent -- the exclusive consideration of
OPE (only $\pi^{\pm}$ with $\Delta I=1$) is not
sufficient and the exchange of vector mesons ($\rho , \omega$)
has to be included. However, the OPE potential is not
negligible due to its long range. The resulting effective $NN$ 
potential is given in Ref.~\cite{DDH80}
\begin{eqnarray}
 V^{pnc}(\vec r,\vec p\,) &=& i\frac{f_{\pi}g_{\pi{NN}}}{2\sqrt{2}M}
    ( \vec{\tau}_1 \times \vec{\tau}_2)_z (\vec{\sigma}_1+
    \vec{\sigma}_2)\cdot 
 \Big[ \vec{p} ,f_{\pi}(r) \Big]  \nonumber \\
 & & -\frac{g_{\rho}}{M} \Big(h_{\rho}^0 \,\vec{\tau}_1\cdot \vec{\tau}_2+
     \frac{h_{\rho}^1}{2} ( \vec{\tau}_1 + \vec{\tau}_2)_z+
     \frac{h_{\rho}^2} {2 \sqrt{6}} (3 \tau_{1,z} \tau_{2,z} -\vec{\tau}_1\cdot 
     \vec{\tau}_2) \Big) \nonumber \\
 & & \times \Big( (\vec{\sigma}_1-\vec{\sigma}_2) \cdot 
     \Big\{ \vec{p},f_{\rho}(r) \Big\} 
   + i(1+\chi_v) \ (\vec{\sigma}_1 \times \vec{\sigma}_2)
    \cdot \Big[ \vec{p},f_{\rho}(r) \Big] \,\Big) 
\nonumber \\
 & & - \frac{g_{\omega}}{M} \Big( h_{\omega}^0 + \frac{h_{\omega}^1}{2} 
     ( \vec{\tau}_1 + \vec{\tau}_2)_z\, \Big) 
\nonumber \\
 & &  \times 
\Big( (\vec{\sigma}_1-\vec{\sigma}_2) \cdot 
    \Big\{ \vec{p},f_{\omega}(r) \Big\} 
    + i(1+\chi_s) \ (\vec{\sigma}_1 \times \vec{\sigma}_2)
    \cdot \Big[ \vec{p},f_{\omega}(r) \Big] \,\Big)
\nonumber \\
 &  & -\frac{g_{\omega} h_{\omega}^1 - g_{\rho} h_{\rho}^1}{2M}\,
      ( \vec{\tau}_1 - \vec{\tau}_2)_z 
      (\vec{\sigma}_1+\vec{\sigma}_2) \cdot 
      \Big\{ \vec{p},f_{\rho}(r) \Big\}\nonumber \\
 & & -i\frac{g_{\rho} h_{\rho}^{\prime 1}}{2M}( \vec{\tau}_1 
     \times \vec{\tau}_2)_z (\vec{\sigma}_1+\vec{\sigma}_2)\cdot 
     \Big[ \vec{p},f_{\rho}(r) \Big],
\label{pv_pot}
\end{eqnarray}
with the usual Yukawa function 
\begin{eqnarray}
f_{\xi}(r) =  \frac{e^{-m_{\xi}r}}{4\pi r} \,, \quad \mbox{for} 
\quad \xi=\pi,\rho,\omega\ .
\end{eqnarray}
Furthermore, $M$ denotes the nucleon mass and $\vec p= \frac{1}{2}(\vec p_1
-\vec p_2)$. Here, seven a priori unknown weak coupling constants appear. 
Specific values for them are given below. 

In order to calculate the opposite parity admixture of the deuteron wave 
function, perturbation theory of first order is sufficient. The admixed 
parity violating component is then given by
\begin{eqnarray}
 \vert \,d_{pnc}^{\,(1)}\, \rangle= -\frac{1}{H_{pc}^{\,(0)} - E} \ 
 \Big(H_{pnc}^{\,(1)}-(E-E_B) \Big) \vert \,d_{pc}^{\,(0)}\, \rangle\, ,
\end{eqnarray}
where $\ \vert \,d_{pc}^{\,(0)}\, \rangle$ is the unperturbed deuteron 
wave function, $H_{pc}^{\,(0)}$ and $H_{pnc}^{\,(1)}$ are the 
unperturbed strong Hamiltonian and the parity violating perturbation, 
respectively, $E_B$ is the unperturbed deuteron binding energy ($=-2.2246$ 
MeV) and $E$ 
the eigenvalue of the perturbed wave function. The difference $(E-E_B)$ 
is of second order and, thus, can safely be ignored.
In the following calculation, the propagator was approximated by the 
free Green function. Applying the potential (\ref{pv_pot}) to the 
unperturbed deuteron wave function yields for the parity violating admixture
\begin{eqnarray}
 \psi_d^{pnc}(\vec{p}\,)&=& \frac{i}{p}\Big(
\tilde{u}_{11}(p)\,\langle \hat p|10;(11)1m_d\rangle
+\tilde{u}_{10}(p)\,\langle \hat p|00;(10)1m_d\rangle\Big)\,,
\end{eqnarray}
where $\tilde{u}_{LS}(p)$ denotes the radial part and 
$\langle \hat p|Tm_T;(LS)JM\rangle$ the isospin, spin and angular momentum 
part of the wave function. In the familiar spectroscopic 
notation, this corresponds to $^3P_1$ und $^1P_1$ states.  
Both parts obey the Pauli principle, i.e., the wave function is 
antisymmetric. The radial functions are given by
\begin{eqnarray}
 \tilde{u}_{11}(p) &=&  -\frac{1}{\pi M\sqrt{3\pi}}\, 
\frac{p}{E_B-\frac{p^2}{M}} \sum_{L=0,2} {(\sqrt{2}\,)}^{-L/2}  \, 
\int dr\,j_1(pr)\nonumber\\
&&
 \bigg\{ \frac {f_{\pi}g_{\pi N N}}{\sqrt{2}}\,e^{-m_{\pi}r} 
 (m_{\pi} + \frac{1}{r}) u_L(r) 
- g_{\rho} h_{\rho}^{\prime 1} e^{-m_{\rho}r}
 (m_{\rho} + \frac{1}{r}) u_L(r) \nonumber \\
&& 
 - (g_{\omega} h_{\omega}^1 - g_{\rho} h_{\rho}^1) 
 e^{-m_{\rho}r}\Big[ (m_{\rho} + (-)^{L/2}\frac{3}{r}) u_L(r) 
 - 2 u_{L}^{\prime}(r) \Big] 
 \bigg\} 
\end{eqnarray}
and
\begin{eqnarray}
  \tilde{u}_{10}(p) &=& \frac{1}{\pi M\sqrt{6\pi}}\, 
 \frac{p}{E_B-\frac{p^2}{M}} \sum_{L=0,2} {(-\sqrt{2}\,)}^{L/2}
  \int dr\,j_1(pr)\nonumber\\
&& \bigg\{ -3g_{\rho} h_{\rho}^0 e^{-m_{\rho}r}
  \Big( [(m_{\rho} + \frac{1}{r})\chi_v +\frac{(-2)^{1+L/2}}{r}]u_L(r) 
 + 2 u_{L}^{\prime}(r) \Big)
\nonumber \\
&&
 + g_{\omega} h_{\omega}^0 e^{-m_{\omega}r}
 \Big([(m_{\omega} + \frac{1}{r})\chi_s +\frac{(-2)^{1+L/2}}{r}]u_L(r) 
 + 2 u_{L}^{\prime}(r) \Big) \bigg\} \,.\label{pnc10}
\end{eqnarray}
According to 
Ref.\ \cite{Hol81}, $h_{\rho}^{\prime 1}$ is comparably small and thus is 
usually neglected. There exist various approaches to give theoretical 
estimations for the remaining coupling constants, e.g. 
\cite{DDH80,GaR74,KKW79,DuZ86}. In Ref.\ \cite{DDH80} a unified treatment 
based on the quark model is performed and we use the estimated ``best values'' 
given there for the weak coupling constants and listed in Table \ref{tab1}.

The radial functions are shown in Fig.\ \ref{deut_wf}. For the 
unperturbed wave function, the parametrization by Machleidt et al.\ 
\cite{MHE87} for the Bonn OBEPQ model has been used. In momentum space, 
the parity violating $P$ waves are much 
more spread out than the normal $S$ and $D$ waves indicating a much shorter 
range in coordinate space. This is particularly pronounced in the 
$^1P_1$-part because the $\pi$ exchange part of $V^{pnc}$ does not 
contribute here (see (\ref{pnc10})). The total $P$-state probability is 
$P_P=2.7 \times 10^{-14}$, which corresponds to an admixture amplitude that is 
of the order of magnitude expected \cite{McK69} and in qualitative agreement 
with \cite{HwH81}. 

For the explicit calculation of the structure functions, we have
chosen the quasifree region where the total momentum of the exchanged boson 
can be transferred completely to one nucleon. In this case, the active 
nucleon is ejected in the forward direction with respect to $\vec q$ while 
the other nucleon acts as a spectator and remains at rest in the lab system. 
The momentum transfer and the relative $np$ final energy are then no 
longer independent. Their dependence defines the quasifree ridge in the 
$E_{np}$-$\vec q^{\,2}$-plane which is characterized by 
$E_{np}/{\mbox{MeV}}\approx 10\,\vec q^{\,2}/{\mbox{fm}}^{-2}$.  
The quasifree kinematics has the advantage that the final state interaction 
and two-body effects from meson exchange currents 
are small \cite{Are82} and thus will be neglected here. 
This means that we evaluate the $T$-matrix elements of (\ref{tgamma}) in the 
Born approximation, taking plane waves as final $np$ scattering states. 
Furthermore, as electromagnetic current we use the one-body 
nucleon current in nonrelativistic reduction to the order 
${\cal{O}}(1/M)$ as is given by
\begin{eqnarray}
\langle N(p_f,s_f)|\rho^{e.m.} (0)|N(p_i,s_i)\rangle & = & 
\chi_{s_f}^{\dagger}\,G_E(q^2)\,\chi_{s_i} \ ,\label{rem}\\
\langle N(p_f,s_f)|\vec{J}^{e.m.}(0)|N(p_i,s_i)\rangle & = & 
\chi_{s_f}^{\dagger}\,\frac{1}{2M} \Big[G_E(q^2)(\vec{p}_f+\vec{p}_i) 
+iG_M(q^2) (\vec{\sigma} \times \vec{q}\,)\Big]\, 
\chi_{s_i} \ ,
\label{jem}
\end{eqnarray} 
where $\chi_s$ denotes a Pauli spinor, $\vec{q}=\vec{p}_f - \vec{p}_i$ 
the momentum transfer, and $G_{E/M}(q^2)$ electric and magnetic Sachs 
form factors, respectively. We use the dipole parametrization with 
$G_{En}$ in the Galster form \cite{GaK71}. 

In view of the restriction to the one-body current, a remark concerning 
current conservation is in order. In principle, one has to include 
in addition also meson exchange current contributions in order to insure 
gauge invariance. In fact, such meson exchange currents are shown to give a 
dominant contribution to the nuclear anapole moment \cite{HaH89} in heavy 
nuclei. But in the case of the deuteron, in particular for the quasifree 
kinematics we expect that such two-body effects are small, similar to the 
parity conserving contribution for which this has been shown to be the case 
\cite{Are82}. But certainly, it remains to be seen what their effect will 
be away from the quasifree case. 

We show in Fig.\ \ref{fga0} for two momentum transfers ($\vec q^{\, 2}=
4\, \mbox{fm}^{-2}$ and $12\, \mbox{fm}^{-2}$, corresponding to 
$E_{np}=40$ MeV and 120 MeV, respectively) the pnc structure functions 
$f_T^{\prime\,\gamma_a}$ and $f_{LT}^{\prime\,\gamma_a}$, which contribute 
to the helicity dependent part of (\ref{diffcross}). The transverse 
structure function $f_T^{\prime\,\gamma_a}$ for polarized electrons is of 
special interest, since its pc equivalent $f_T^{\prime\,\gamma_v}$ 
is absent. As shown in Fig.~\ref{fga0}, this structure function has a
pronounced minimum at small forward angles and a second peak at
backward angles. There is no fundamental change for the higher momentum 
transfer (see right panel of Fig.\ \ref{fga0}). 
All in all, this structure function is suppressed by a factor
of $10^{-8}$ compared to the pc structure function
$f^{\gamma_v}_T$ \cite{ArL92}. The structure function 
$f_{LT}^{\prime\,\gamma_a}$
shows a sharp maximum at small forward angles that narrows for the
higher momentum transfer but the height remains unchanged. 
$f_{LT}^{\prime\,\gamma_a}$ is suppressed by roughly $10^{-5}$
compared to the parity conserving structure function
$f^{\gamma_v}_{LT}$. 

With respect to target orientation, we note that in the parity 
conserving case, the structure functions 
$f_L^{1M,\gamma_v}$, $f_T^{1M,\gamma_v},
f_{LT}^{1M,\gamma_v}$, $f_{TT}^{1M,\gamma_v}$, $f_{LT}^{\prime\,2M,\gamma_v}$
and $f_{T}^{\prime\,2M,\gamma_v}$ vanish in Born approximation, as
can be seen from (\ref{ffirst}) through (\ref{flast}), because in this case the 
corresponding $v$'s of (\ref{vlam}) are real quantities. The nonvanishing
parity violating counterparts are shown in Figs.~\ref{fga1} and 
\ref{fga2} for the projection $M=0$. In Fig.\ \ref{fga1} we present 
the pnc structure functions $f_\alpha^{10,\gamma_a}$ for vector polarized 
deuterons. Except for $f_{TT}^{10,\gamma_a}$ they exhibit a 
typical forward/backward peaking. Finally, the pnc structure functions for 
a tensor polarized deuteron target are shown in Fig.\ \ref{fga2}. The 
largest pnc structure functions for an oriented target are in decreasing 
order $f_{LT}^{\prime\,20,\gamma_a}$, $f_{L}^{10,\gamma_a}$ and 
$f_{T}^{10,\gamma_a}$. In general one sees that for a polarized target, 
the relative order of magnitude between pc and pnc structure functions
varies from $10^{-4}$ to $10^{-7}$. 


\section{Electroweak interference}
\label{ewint}

As we have seen in Sect.\ \ref{formew}, the presence of the 
electroweak interference leads to additional structure functions in the 
expression (\ref{s-gam-2}) for the differential
cross section. One should keep in mind that the terms proportional to 
$a_a$ result from the interference of the leptonic axial current with 
the weak hadronic vector current, whereas the structure
functions $f_{\alpha}^{(\prime) IM\pm,\,Z_a}$ are a manifestation of the
hadronic axial current and thus have a counterpart in 
$f_{\alpha}^{(\prime) IM\pm,\,\gamma_a}$ resulting from pnc deuteron 
components. 

For the explicit calculation of the structure functions, the same
assumptions are made as in Sect.\ \ref{hadrpv}, i.e., quasifree kinematics and
Born approximation.  The weak neutral nucleon current consists of a vector 
and an axial current. The vector current has an analogous form as the e.m. 
current in Eqs.\ (\ref{rem}) and (\ref{jem}) but with weak Sachs form 
factors $\tilde G_{E/M}(q^2)$ which are given by 
\begin{eqnarray}
\tilde G_{E/M}(q^2)&=& \frac{1}{2}(1-2\sin^2\theta_W)\tau_3G_{E/M}^{(1)}(q^2)
  - \sin^2\theta_W G_{E/M}^{(0)}(q^2)-\frac{1}{4}G_{E/M}^{(s)}(q^2)\,,
\end{eqnarray}
where for $X=E,\, M$
\begin{eqnarray}
G_{X}^{(0/1)}(q^2)&=& \frac{1}{2}\Big(G_{X,p}(q^2)\pm G_{X,n}(q^2)\Big)
\label{sachsiso}
\end{eqnarray}
denote the isoscalar and isovector e.m.\ nucleon Sachs form factors and 
$G_{E/M}^{(s)}(q^2)$ the strangeness form factor, i.e., the contribution of 
the $s\bar s$ density of the nucleon. 

The axial current of the nucleon in nonrelativistic reduction is given by
\begin{eqnarray}
\langle N(p_f,s_f)|\rho^a(0)|N(p_i,s_i)\rangle & = & \chi_{s_f}^{\dagger}\,
\Big[\frac{G_A(q^2)}{2M} \vec{\sigma}\cdot (\vec{p}_i+\vec{p}_f) \Big]\,
\chi_{s_i} \\
\langle N(p_f,s_f)|\vec{J}^{\,a}(0)|N(p_i,s_i)\rangle & = & 
-\chi_{s_f}^{\dagger}\,\Big[G_A(q^2)\vec{\sigma} \Big]\, \chi_{s_i} \ ,
\end{eqnarray}
with the weak axial nucleon form factor 
\begin{eqnarray}
G_A(q^2)&=&-\frac{1}{2}\tau_3 G_A^{(1)}(q^2) + \frac{1}{4}G_A^{(s)}(q^2) \,,
\label{axialff}
\end{eqnarray}
where $G_A^{(1)}(q^2)$ is defined in (\ref{sachsiso}) for $X=A$. 
For the axial form factor a dipole parametrization with mass parameter $M_a =
1.032$~GeV is assumed with $G_A^{(1)}(0)=1.262$ and $G_A^{(s)}(0)=.38$ 
determined from 
neutron beta decay and neutrino scattering data, respectively \cite{BMK91}.
For the isovector part of the weak vector form factors we use a dipole fit, 
while for the isoscalar part and the strange form factors we use the fit 
of Jaffe et al.\ \cite{Jaf89}. We would like to remark that (i) 
the Eq.\ (\ref{axialff}) is valid at tree level only, because electroweak 
radiative corrections generate additional isoscalar axial vector 
components, and (ii) the fit of \cite{Jaf89} is not realistic at high 
momentum transfers \cite{MuD92}. However, in view of the fact, that we do 
not consider very high momentum transfers and that we are mainly interested 
in the relative size of electroweak and hadronic parity violation, we do 
not consider these shortcomings too serious. 

Because of the Born approximation, resulting in real $v$'s, some of the 
structure functions vanish identically according to (\ref{fcv})-(\ref{fca}). 
The remaining pnc structure functions from $Z$ exchange 
without target polarization are shown in Figs.\ \ref{fzvdiag} through 
\ref{fza} for the same kinematics as in Sect.\ \ref{hadrpv}. Those for 
the weak hadronic vector current are shown in Figs.\ \ref{fzvdiag} 
and \ref{fzvintf}, while Fig.~\ref{fza} refers to the hadronic 
axial current. Here, all pnc structure functions were multiplied 
with the appropriate factor $a_v$ or $a_a$ as they appear in the helicity 
dependent part of the differential cross section (see (\ref{diffcross})) 
in order to represent the appropriate order of magnitude. 
As has been mentioned, the pnc 
structure functions from the hadronic vector current are related to
the leptonic axial current and thus have no equivalent in the case of
$P$-wave admixture in the deuteron. 
Their size is about $10^{-4}-10^{-7}$ relative to the pc structure functions. 
Except for $f_{TT}^{Z_v}$ at $\vec q^{\,2}=4\,\mbox{fm}^{-2}$, 
all pnc structure functions exhibit enhanced maxima or minima at forward 
and backward angles, typical for quasifree kinematics. They show 
furthermore a stronger dependence on $\vec q^{\,2}$ than the corresponding 
structure functions for the pnc deuteron components, because for the latter 
the transition current density is much shorter ranged. 

The structure function $f_T^{\prime\,Z_a}$ in Fig.~\ref{fza} is in absolute
value about three orders of magnitude higher than its equivalent 
$f_T^{\prime\,\gamma_a}$ due to the $P$-wave admixture. The peaking at 
forward and backward angles is narrower and for the forward angles of 
opposite sign (cf.\ Fig.~\ref{fga0}). This is not true for
$f_{LT}^{\prime\,Z_a}$ which is of the same order of magnitude, but 
slightly smaller than the corresponding $P$-wave structure function 
$f_{LT}^{\prime\,\gamma_a}$ in Fig.\ \ref{fga0}. It shows a similar 
behaviour at forward angles, whereas its minimum at backward angles is much 
more pronounced.


\section{Inclusive longitudinal asymmetry} 
\label{asym}

The parity violating observable that is measured in inclusive electron 
scattering off an unpolarized deuteron target is the longitudinal 
asymmetry
\begin{eqnarray}
 {\cal A} & = & \frac{1}{2h\Sigma_0}(\Sigma_+-\Sigma_-) \, ,
\end{eqnarray}
where 
\begin{eqnarray}
\Sigma_0&=&\frac{1}{2}\Big(\Sigma_+ +\Sigma_-\Big)\,.
\end{eqnarray}
With the inclusive differential cross section  from (\ref{inclusive}), 
$\Sigma_{\pm} = \Sigma(\pm h)$ for $h > 0$, neglecting the tiny 
contributions from $Z$ exchange, one finds
\begin{eqnarray}
\Sigma_0 &=& c\Big(\rho_L F_L^{\gamma_v}+\rho_T F_T^{\gamma_v}\Big)\,.
\end{eqnarray}
According to (\ref{inclusive}), one may split the asymmetry into the 
contributions from hadronic and electroweak parity violation, namely 
\begin{eqnarray}
{\cal A}= {\cal A}^{\gamma_a}+ {\cal A}^{Z}\,,
\end{eqnarray}
where
\begin{eqnarray}
 {\cal A}^{\gamma_a}& = & -\frac{\rho_T^{\prime}F_T^{\prime\,{\gamma_a}}}
{\rho_L F_L^{\gamma_v}+\rho_T F_T^{\gamma_v}}\,, \\ 
 {\cal A}^{Z} & = & \frac{\rho_L a_a F_L^{Z_v} + 
\rho_T a_a F_T^{Z_v}-\rho_T^{\prime}a_v F_T^{\prime\,Z_a}}
{\rho_L F_L^{\gamma_v}+\rho_T F_T^{\gamma_v}} \nonumber\\
 & = & \frac{G_F}{2\sqrt{2}}\frac{q_\mu^2}{\pi \alpha}\,\,
\frac{\rho_L F_L^{Z_v} +
\rho_T F_T^{Z_v}+\rho_T^{\prime}(1-4\sin^2\theta_W)F_T^{\prime\,Z_a}}
{\rho_L F_L^{\gamma_v}+\rho_T F_T^{\gamma_v}}\,.
\end{eqnarray}

The contributions of the different inclusive form factors are exhibited in
Fig.~\ref{pncff} along the quasifree ridge. The largest form factor is the 
transverse one for the weak hadronic vector current $a_aF_T^{Z_v}$. The 
other two electroweak form factors, $a_aF_L^{Z_v}$  and 
$a_v F_T^{\prime\,Z_a}$, reach only one third in magnitude but have the 
opposite sign than $a_aF_T^{Z_v}$. Three orders of magnitude smaller is 
$F_T^{\prime \,\gamma_a}$, the only form factor from the pnc 
deuteron components. The resulting longitudinal beam asymmetries 
${\cal A}^{\gamma_a}$ and ${\cal A}^{Z}$ are shown in 
Fig.~\ref{longasy}. Since for a given momentum transfer $q^{lab}$ 
the asymmetry depends also on the electron kinematics through the lepton 
density matrices, we have chosen two laboratory scattering angles, one at 
a more forward direction ($\theta_e=35^\circ$) and one backward angle 
($\theta_e=170^\circ$). One readily notes that the 
beam asymmetry ${\cal A}^{\gamma_a}$ due to the $P$-wave in the deuteron 
varies strongly with the scattering angle. Indeed, it is relatively more 
suppressed for backward scattering than ${\cal A}^{Z}$. Furthermore, it is 
apparent that ${\cal A}^{\gamma_a}$ is negligible compared to ${\cal A}^{Z}
$ over the whole range of momentum transfers considered here. This means 
that even for the low momentum transfers of the SAMPLE experiment \cite{BeA96} 
the contributions from the pnc deuteron components can be neglected. 

For the beam asymmetry due to electroweak interference, the dominant
contribution comes from the term proportional to $a_aF_T^{Z_v}$, but the
dependence on the electron scattering angle is mainly a result of the term
proportional to $a_aF_L^{Z_v}$. In order to compare 
our results on the asymmetry ${\cal A}^{Z}$ from electroweak interference 
with those reported in \cite{HPD92}, we show in Fig.~\ref{longasylog} the 
asymmetries on a logarithmic scale. For ${\cal A}^{Z}$ we find very good 
agreement with their results for the quasifree case with plane wave Born
approximation. 


\section{Summary and conclusions }
\label{concl}

In this study two mechanisms of the weak interaction have been
investigated that introduce parity violation in electrodisintegration 
of the deuteron, namely electroweak $\gamma$-$Z$ interference and parity 
violating components in the nuclear wave function. We have derived the 
formal expressions for all observables including beam polarization and target 
orientation in terms of pc and pnc structure functions. For the explicit 
calculation, we have concentrated on the quasifree region where final
state interaction, meson exchange currents and isobar configurations
could be neglected. 

In the hadronic part a parity violating $NN$ potential was considered 
which led to an abnormal parity admixture in the deuteron wave function. 
The $P$-wave component in the deuteron wave function consists of two 
parts, a spin singlet and a spin triplet. The total $P$-state probability 
is very small $(2.7\times 10^{-12}\%)$. Compared to the pc structure 
functions, the pnc structure functions are suppressed by orders 
$10^{-4}-10^{-7}$. The resulting parity violating form factors are three 
orders of magnitude smaller than the ones from $Z$-exchange. For this 
reason one can safely neglect the contribution of the $P$-wave admixture 
in the deuteron to the beam asymmetry for longitudinally polarized 
electrons in the quasifree domain and need only to consider the dominant 
contribution from $Z$-exchange. 

As already mentioned, there is a great deal of interest in experiments to
determine strange quark contributions to hadronic matrix elements.
Several experiments are underway or planned, e.g., at MIT-Bates
\cite{BeA96}, Mainz \cite{Har93} and CEBAF \cite{FiS91}. In connection
with this study, the SAMPLE experiment at MIT-Bates \cite{BeA96} is of
special interest, since it measures the strange magnetic form factor
$G_M^{(s)}$ at quite low momentum transfer as determined in parity violating
electron scattering off hydrogen and deuterium. Even here, we have not 
found significant effects from parity violation in the hadronic sector. 

In order to study parity violation which originates from the hadronic sector 
in electromagnetic deuteron break-up, one has to go away from the quasifree 
kinematics to lower momentum transfer where it may become more comparable 
in size to the contribution from electroweak interference. But then one has 
to consider also pnc components in the final state and the contribution from 
meson exchange currents. In this respect, 
we consider this study as starting point for further investigations, in 
particular with respect to the role of final state interaction, meson exchange 
currents and isobar configurations.


\renewcommand{\theequation}{A.\arabic{equation}}
\setcounter{equation}{0}
\section*{Appendix: Polarization observables and structure functions} 
In this appendix we will give explicit expressions for the various polarization 
observables in terms of corresponding structure functions. Denoting a 
general polarization observable by $X$ as listed in Table \ref{tab2} 
one finds 
\begin{eqnarray}
 S_X(h,P_1^d,P_2^d) &=& P_X
\frac{d^{3}\sigma^{\gamma +Z}}{dk_2^{ lab} d \Omega_{k_2}^{ lab} 
d\Omega_{np}^{ c.m.}} \nonumber\\
&=&
 \bar S_X(\gamma_v;P_1^d,P_2^d)+\bar S_X(\gamma_a;P_1^d,P_2^d)\nonumber\\
&&
 +a_v\Big(\bar S_X(Z_v;P_1^d,P_2^d)+\bar S_X(Z_a;P_1^d,P_2^d)\Big)\nonumber\\
&&
 +a_a\Big(\bar S_X'(Z_v;P_1^d,P_2^d)+\bar S_X'(Z_a;P_1^d,P_2^d)\Big)
\nonumber\\
&&
+h\Big[
 \bar S_X'(\gamma_v;P_1^d,P_2^d)+\bar S_X'(\gamma_a;P_1^d,P_2^d)\nonumber\\
&&
 +a_v\Big(\bar S_X'(Z_v;P_1^d,P_2^d)+\bar S_X'(Z_a;P_1^d,P_2^d)\Big)\nonumber\\
&&
 +a_a\Big(\bar S_X(Z_v;P_1^d,P_2^d)+\bar S_X(Z_a;P_1^d,P_2^d)\Big)\Big]\,,
\label{x-gam-2}
\end{eqnarray}
where, in analogy to (\ref{obsfins}), we have defined for 
$C\in {\cal S}_v\cup {\cal S}_a$
\begin{eqnarray}
\bar S_X(C;\,P_1^d,P_2^d)
&=&
 c \sum _{I=0}^2 P_I^d \sum _{M=0}^I
 \Bigl\{  (\rho _L f_L^{IM,C}(X) + \rho_T f_T^{IM,C}(X) +
 \rho_{LT} {f}_{LT}^{IM+,C}(X) \cos \phi \nonumber\\
& &
 + \rho _{TT} {f}_{TT}^{IM+,C}(X) \cos2 \phi)
 \cos (M\tilde{\phi}-\bar\delta_{I}^{C,X} {\pi \over 2})\nonumber \\
& & 
-(\rho_{LT} {f}_{LT}^{IM-,C}(X) \sin \phi
+ \rho _{TT} {f}_{TT}^{IM-,C}(X) \sin2 \phi)
\sin (M\tilde{\phi}-\bar\delta_{I}^{C,X} {\pi \over 2})
\Big\} \nonumber \\
& &
\times d_{M0}^I(\theta_d)\,,\label{xobsfin}\\
\bar S_X'(C;\,P_1^d,P_2^d)
&=&
 c \sum _{I=0}^2 P_I^d \sum _{M=0}^I
 \Bigl\{ (\rho'_T f_T^{\prime IM,C}(X)
 + \rho '_{LT} {f}_{LT}^{\prime IM-,C}(X) \cos \phi ) 
 \sin (M\tilde{\phi}-\bar\delta_{I}^{C,X} {\pi \over 2}) \nonumber\\
& &
 \qquad + \rho '_{LT} {f}_{LT}^{\prime IM+,C}(X) \sin \phi
 \cos (M\tilde{\phi}-\bar\delta_{I}^{C,X} {\pi \over 2})
 \Big\} d_{M0}^I(\theta_d)\,,\label{xobsfins}
\end{eqnarray}
introducing as generalization of (\ref{deltaic}) 
\begin{equation}
\bar\delta_{I}^{C,X}:=\left\{\matrix{(\delta_{X,B}-\delta_{I1})^2 & 
\mbox{for}\; C\in {\cal S}_v \cr (\delta_{X,A}-\delta_{I1})^2 & 
\mbox{for}\; C\in {\cal S}_a \cr} \right\}\,.
\end{equation}

The structure functions for a general observable $X$ are given by 
\begin{eqnarray}
f_{L}^{IM,C}(X)&=&\frac{2}{1+\delta_{M0}}\Re\Big[i^{\bar \delta^{C,X}_I}
{\cal U}^{00 I M,C}_{X}\Big]\,,\\
f_{T}^{IM,C}(X)&=&\frac{4}{1+\delta_{M0}}\Re\Big[i^{\bar \delta^{C,X}_I}
{\cal U}^{11 I M,C}_{X}\Big]\,,\\
f_{LT}^{IM\pm,C}(X)&=&\frac{4}{1+\delta_{M0}}\Re\Big[i^{\bar \delta^{C,X}_I}
\left({\cal U}^{01 I M,C}_{X}\pm(-)^{I+M+\delta_{X,\,B}+\delta_C}
{\cal U}^{01 I -M,C}_{X}\right)\Big]\,,\\
f_{TT}^{IM\pm,C}(X)&=&\frac{2}{1+\delta_{M0}}\Re\Big[i^{\bar \delta^{C,X}_I}
\left({\cal U}^{-11 I M,C}_{X}\pm(-)^{I+M+\delta_{X,\,B}+\delta_C}
{\cal U}^{-11 I -M,C}_{X}\right)\Big]\,,\\
f_{T}^{\prime IM,C}(X)&=&\frac{4}{1+\delta_{M0}}\Re\Big[i^{1+\bar 
\delta^{C,X}_I}
{\cal U}^{11 I M,C}_{X}\Big]\,,\\
f_{LT}^{\prime IM\pm,C}(X)&=&\frac{4}{1+\delta_{M0}}\Re\Big[i^{1+\bar 
\delta^{C,X}_I}
\left({\cal U}^{01 I M,C}_{X}\pm(-)^{I+M+\delta_{X,\,B}+\delta_C}
{\cal U}^{01 I -M,C}_{X}\right)\Big]\,,
\end{eqnarray}
where the ${\cal U}^{\lambda'\lambda I M,C}_X$ are given in terms of 
the quantities
\begin{eqnarray}
u^{s's S\sigma,C}_{\lambda'\lambda I M}=
\frac{\hat I \sqrt 3}{1+\delta_{C,\gamma_v}}\sum_{m_s' m_s m'm}&&
(-)^{1-m+s'-m_s'}\left( \matrix {1&1&I \cr m'&-m&M \cr} \right)
\left( \matrix {s'&s&S \cr m_s'&-m_s&-\sigma  \cr} \right)
\nonumber\\
&&\times\Big[(t^{\gamma_v}_{s'm_s' \lambda m'}(\theta) )^\ast
t_{sm_s \mu m}^C(\theta)+(t^{C}_{s'm_s' \lambda m'}(\theta))^\ast
t_{sm_s \mu m}^{\gamma_v}(\theta)\Big]
\,,\label{defu}
\end{eqnarray}
by the following expressions:\\
(i) differential cross section ($X=1$) 
\begin{eqnarray}
{\cal U}^{\lambda'\lambda I M,C}_{1}&=& 
\sum_s\hat s u_{\lambda'\lambda IM}^{ss00,C}\,,
\end{eqnarray}
(ii) single nucleon polarization
\begin{eqnarray}
{\cal U}^{\lambda'\lambda I M,C}_{x(1)}&=&-\sqrt{\frac{3}{2} }
\Big(V_{\lambda'\lambda I M}^{1011,C}-V_{\lambda'\lambda I M}^{101-1,C}
\Big)\,,\\
{\cal U}^{\lambda'\lambda I M,C}_{x(2)}&=&-\sqrt{\frac{3}{2} }
\Big(V_{\lambda'\lambda I M}^{0111,C}-V_{\lambda'\lambda I M}^{011-1,C}
\Big)\,,\\
{\cal U}^{\lambda'\lambda I M,C}_{y(1)}&=&i\sqrt{\frac{3}{2} }
\Big(V_{\lambda'\lambda I M}^{1011,C}+V_{\lambda'\lambda I M}^{101-1,C}
\Big)\,,\\
{\cal U}^{\lambda'\lambda I M,C}_{y(2)}&=&i\sqrt{\frac{3}{2} }
\Big(V_{\lambda'\lambda I M}^{0111,C}+V_{\lambda'\lambda I M}^{011-1,C}
\Big)\,,\\
{\cal U}^{\lambda'\lambda I M,C}_{z(1)}&=& \sqrt{3}
V_{\lambda'\lambda I M}^{1010,C}\,,\\
{\cal U}^{\lambda'\lambda I M,C}_{z(2)}&=& \sqrt{3}
V_{\lambda'\lambda I M}^{0110,C}\,,
\end{eqnarray}
where
\begin{eqnarray}
V_{\lambda'\lambda I M}^{1 0 1\sigma,C}&=&\sqrt{2}
\sum_{s's}(-)^{s}\hat s'\hat s \left\{ \matrix {s'&s &1 \cr
\frac{1}{2}&\frac{1}{2}&\frac{1}{2} \cr } \right\}
u^{s's 1\sigma,C}_{\lambda'\lambda I M}\,,\\
V_{\lambda'\lambda I M}^{0 1 1\sigma,C}&=&\sqrt{2}
\sum_{s's}(-)^{s'}\hat s'\hat s \left\{ \matrix {s'&s &1 \cr
\frac{1}{2}&\frac{1}{2}&\frac{1}{2} \cr } \right\}
u^{s's 1\sigma,C}_{\lambda'\lambda I M}\,,
\end{eqnarray}
(iii) double nucleon polarization
\begin{eqnarray}
{\cal U}^{\lambda'\lambda I M,C}_{xx/yy}&=&-\sqrt{3}\Big[
V_{\lambda'\lambda I M}^{1 1 00,C}+\frac{1}{\sqrt{2}}
V_{\lambda'\lambda I M}^{1 1 20,C}\mp\frac{\sqrt{3}}{2}
\Big(V_{\lambda'\lambda I M}^{1 1 22,C}+V_{\lambda'\lambda I M}^{1 1 2-2,C}
\Big)\Big]\,,\\
{\cal U}^{\lambda'\lambda I M,C}_{zz}&=&-\sqrt{3}\Big[
V_{\lambda'\lambda I M}^{1 1 00,C}-\sqrt{2}
V_{\lambda'\lambda I M}^{1 1 20,C}\Big]\,,\\
{\cal U}^{\lambda'\lambda I M,C}_{xy/yx}&=&-\frac{3i}{2}\Big[\pm\sqrt{2}
V_{\lambda'\lambda I M}^{1 1 10,C}+
\Big(V_{\lambda'\lambda I M}^{1 1 22,C}-V_{\lambda'\lambda I M}^{1 1 2-2,C}
\Big)\Big]\,,\\
{\cal U}^{\lambda'\lambda I M,C}_{xz/zx}&=&-\frac{3}{2}\Big[\pm
\Big(V_{\lambda'\lambda I M}^{1 1 11,C}+V_{\lambda'\lambda I M}^{1 1 1
-1,C}\Big)+
\Big(V_{\lambda'\lambda I M}^{1 1 21,C}-V_{\lambda'\lambda I M}^{1 1 2-1,C}
\Big)\Big]\,,\\
{\cal U}^{\lambda'\lambda I M,C}_{yz/zy}&=&\frac{3i}{2}\Big[\pm
\Big(V_{\lambda'\lambda I M}^{1 1 11,C}-V_{\lambda'\lambda I M}^{1 1 1-1,C}
\Big)+
\Big(V_{\lambda'\lambda I M}^{1 1 21,C}+V_{\lambda'\lambda I M}^{1 1 2-1,C}
\Big)\Big]\,.
\end{eqnarray}
where the quantities $V_{\lambda'\lambda I M}^{\tau'\tau S\sigma,C}$ are given 
by
\begin{eqnarray}
V_{\lambda'\lambda I M}^{\tau'\tau S\sigma,C}&=&2\hat S
\sum_{s's}\hat s'\hat s
\left\{ \matrix {\frac{1}{2}&\frac{1}{2}&\tau'\cr
\frac{1}{2}&\frac{1}{2}&\tau \cr s'&s &S \cr} \right\}
u^{s's S\sigma,C}_{\lambda'\lambda I M}\,.
\label{defv}
\end{eqnarray}

\begin{figure}
\centerline{\psfig{figure=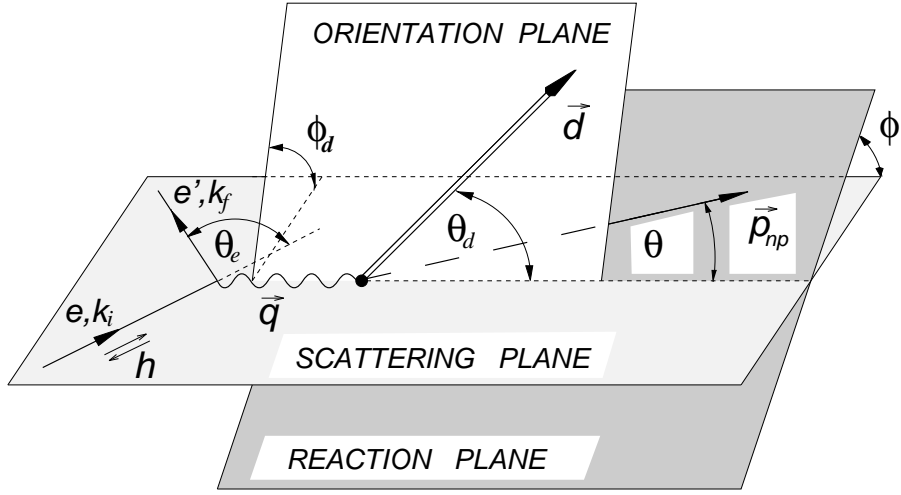,width=12cm,angle=0}}
\vspace*{.5cm}
\caption{
Geometry of exclusive electroweak deuteron disintegration with
polarized electrons and an oriented deuteron target. The relative $np$ 
momentum,
denoted by ${\vec p}_{np}$, is characterized by angles $\theta=\theta_{np}$
and $\phi=\phi_{np}$ where the deuteron orientation axis, denoted by
$\vec d$, is specified by angles $\theta_d$ and $\phi_d$.
}
\label{kine}
\end{figure}

\begin{figure}
\centerline{\psfig{figure=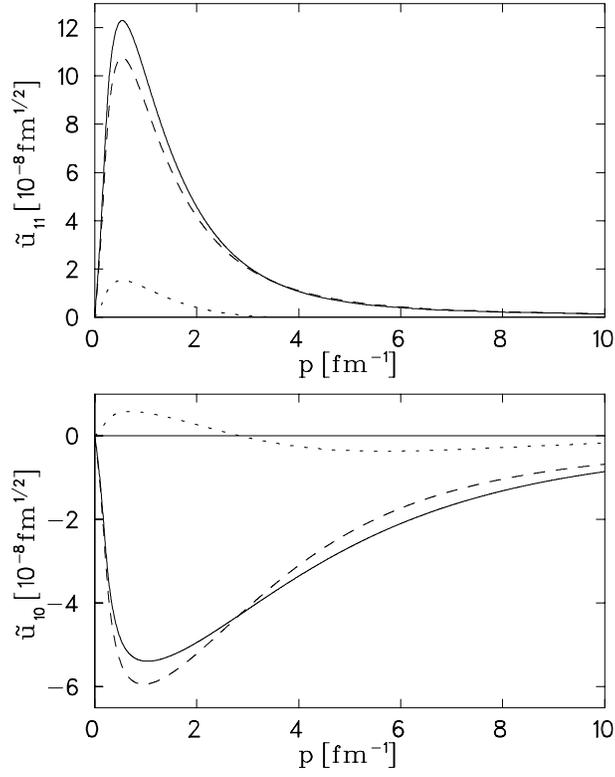,width=8cm,angle=0}}
\vspace*{.5cm}
\caption{
Radial parts of pnc $P$ wave components (full) of the deuteron wave function 
in momentum space. Separately shown are the contributions from $S$ (dashed) 
and $D$ (dotted) components. 
}
\label{deut_wf}
\end{figure}

\begin{figure}
\centerline{\psfig{figure=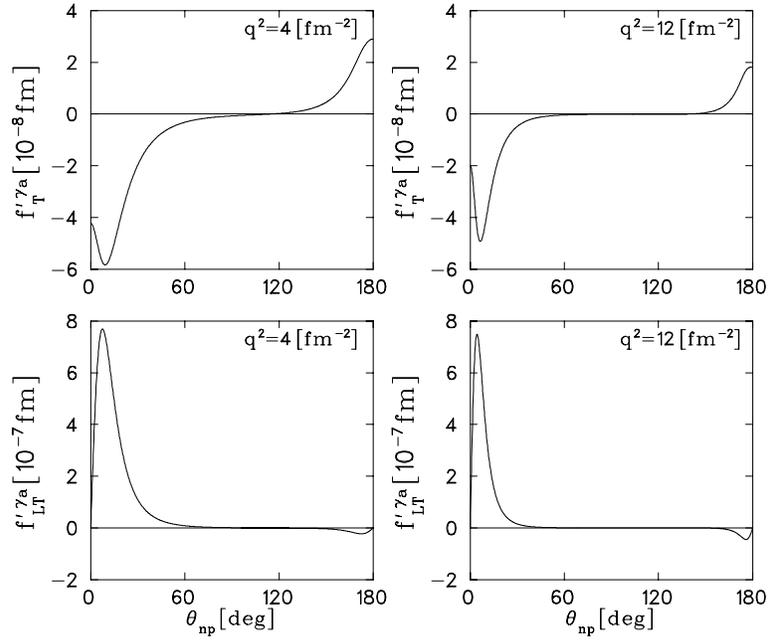,width=10cm,angle=0}}
\vspace*{.5cm}
\caption{
Pnc structure functions from $P$-wave deuteron components for 
unoriented target at $\vec q^{\, 2}=4\,\mbox{and}\,12\,\mbox{fm}^{-2}$. 
}
\label{fga0}
\end{figure}

\begin{figure}
\centerline{\psfig{figure=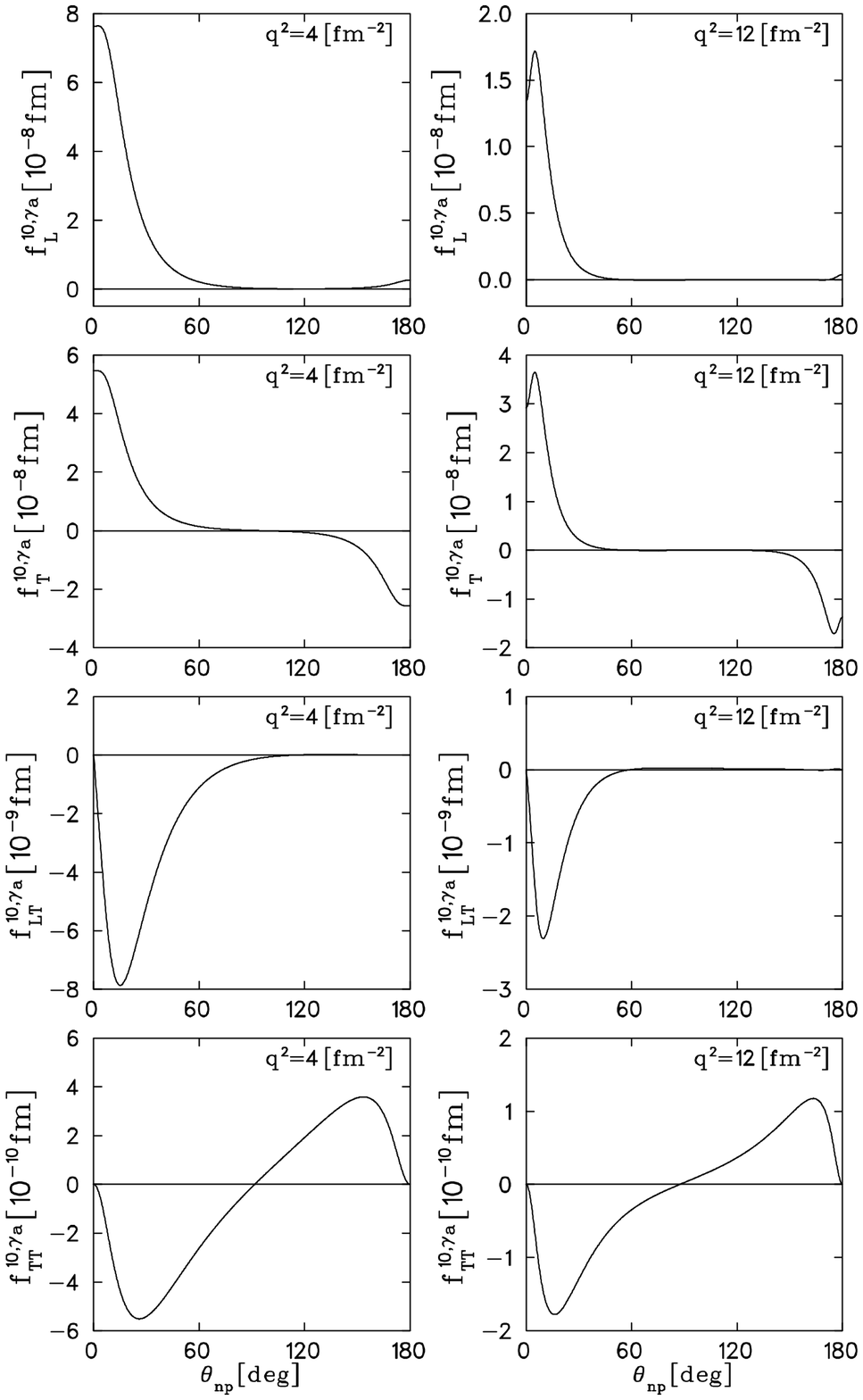,width=12cm,angle=0}}
\vspace*{.5cm}
\caption{
Pnc structure functions from $P$-wave deuteron components for vector 
polarized target at $\vec q^{\, 2}=4\,\mbox{and}\,12\,\mbox{fm}^{-2}$.
}
\label{fga1}
\end{figure}

\begin{figure}
\centerline{\psfig{figure=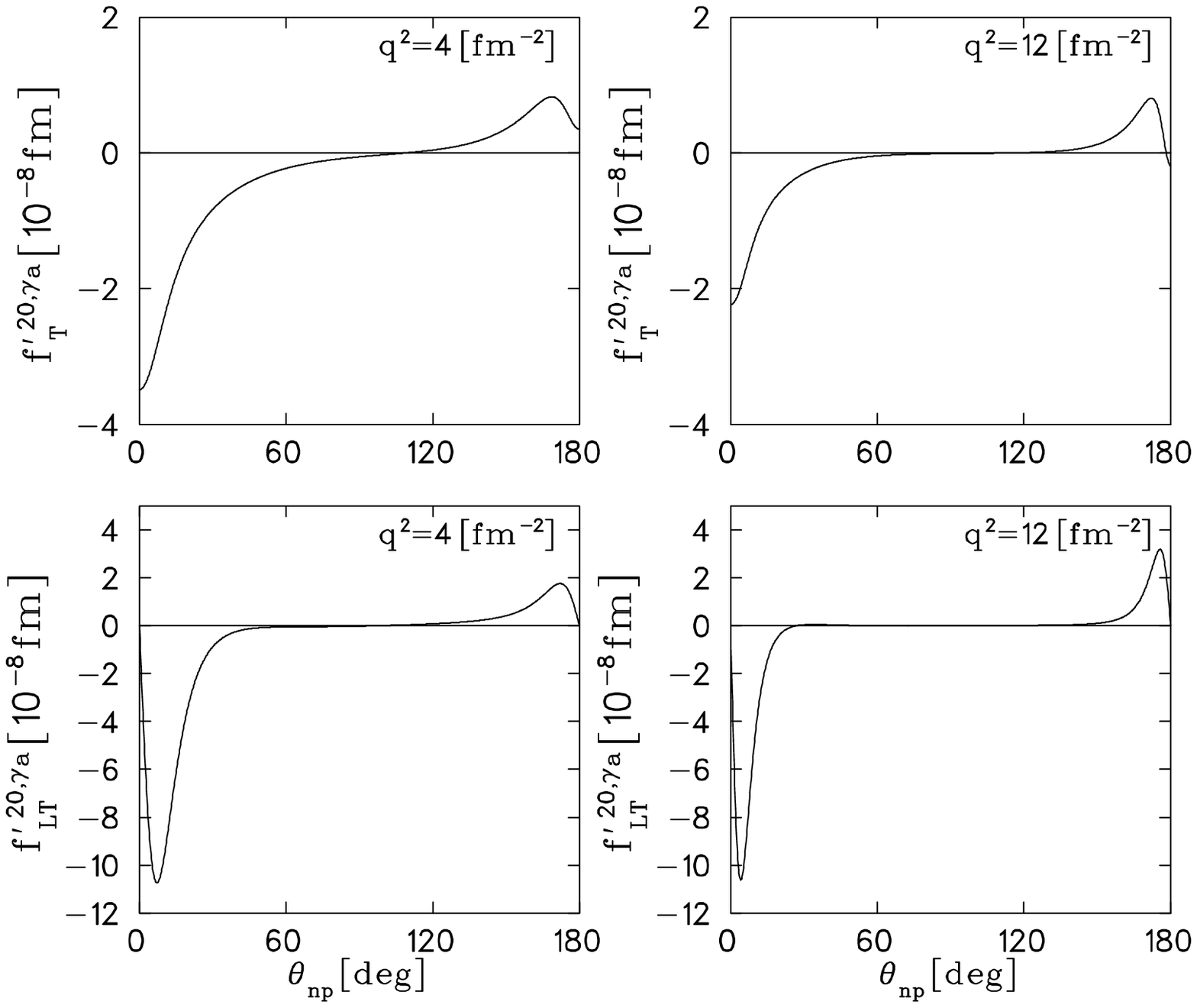,width=10cm,angle=0}}
\vspace*{.5cm}
\caption{
Pnc structure functions from $P$-wave deuteron components for 
tensor polarized target at $\vec q^{\, 2}=4\,\mbox{and}\,12\,\mbox{fm}^{-2}$.
}
\label{fga2}
\end{figure}

\begin{figure}
\centerline{\psfig{figure=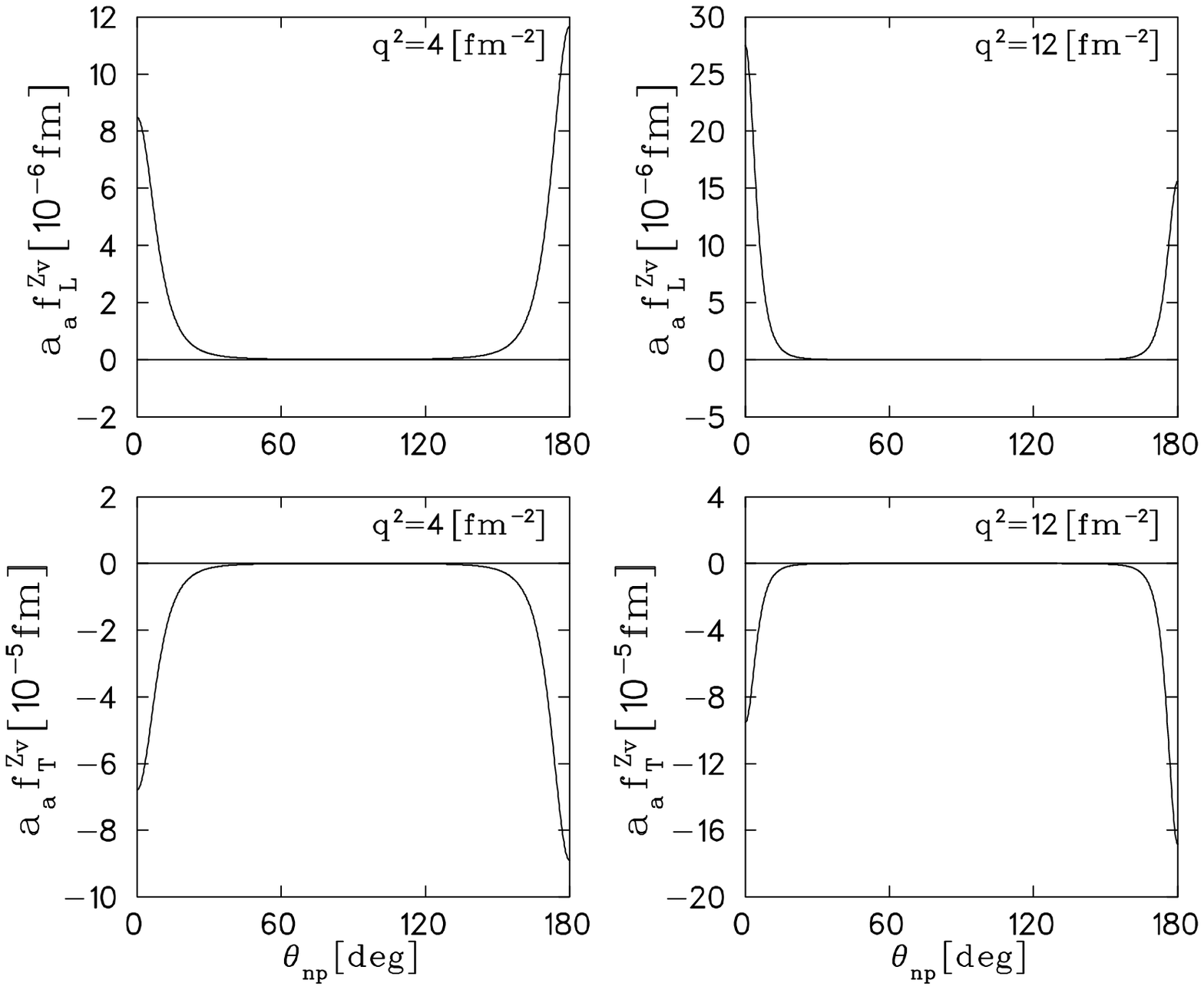,width=10cm,angle=0}}
\vspace*{.5cm}
\caption{
Longitudinal and transverse pnc structure functions from the weak hadronic 
vector current at $\vec q^{\, 2}=4\,\mbox{and}\,12\,\mbox{fm}^{-2}$.
}
\label{fzvdiag}
\end{figure}

\begin{figure}
\centerline{\psfig{figure=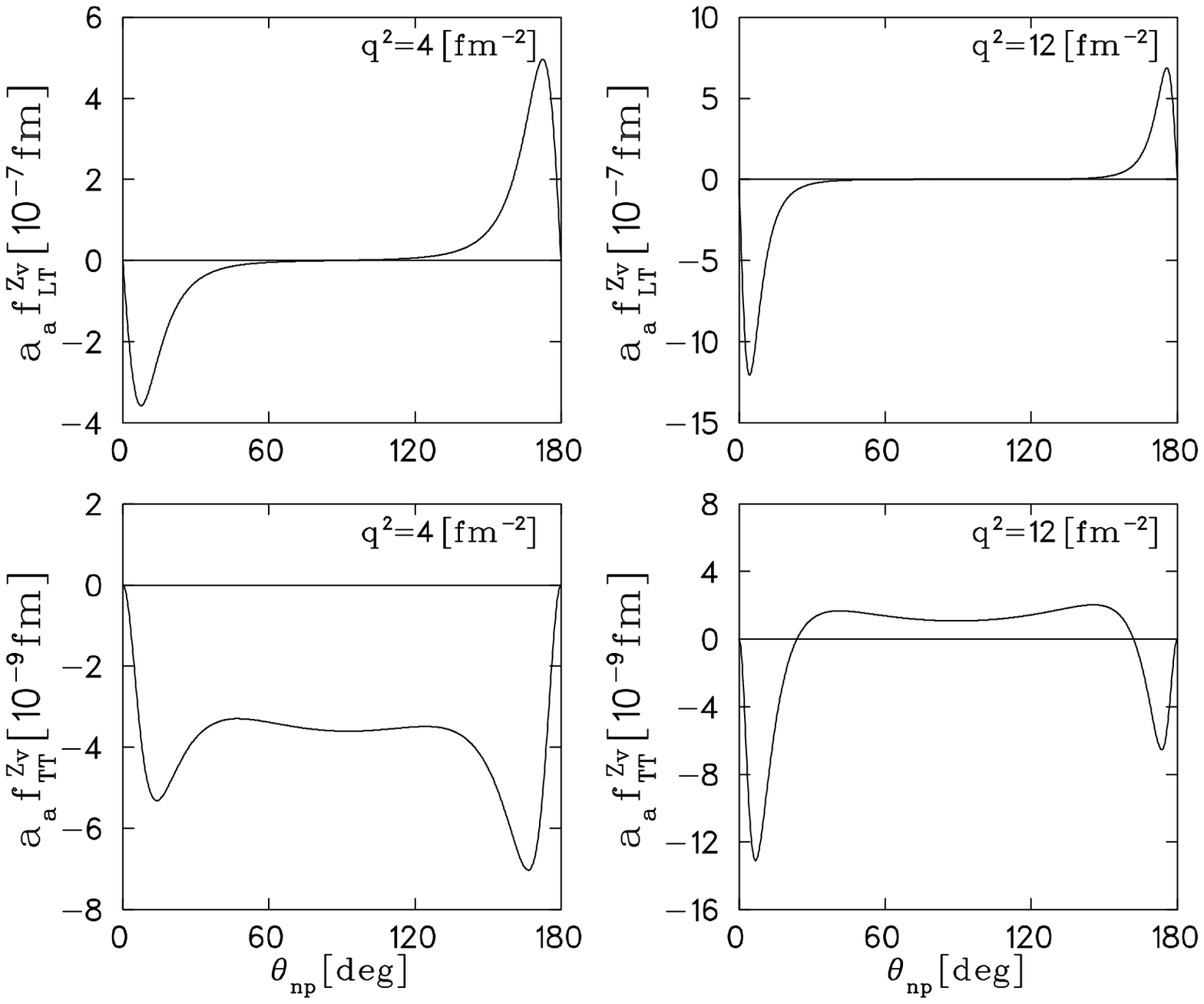,width=10cm,angle=0}}
\vspace*{.5cm}
\caption{
Longitudinal-transverse and transverse-transverse pnc structure functions 
from the weak hadronic
vector current at $\vec q^{\, 2}=4\,\mbox{and}\,12\,\mbox{fm}^{-2}$.
}
\label{fzvintf}
\end{figure}

\begin{figure}
\centerline{\psfig{figure=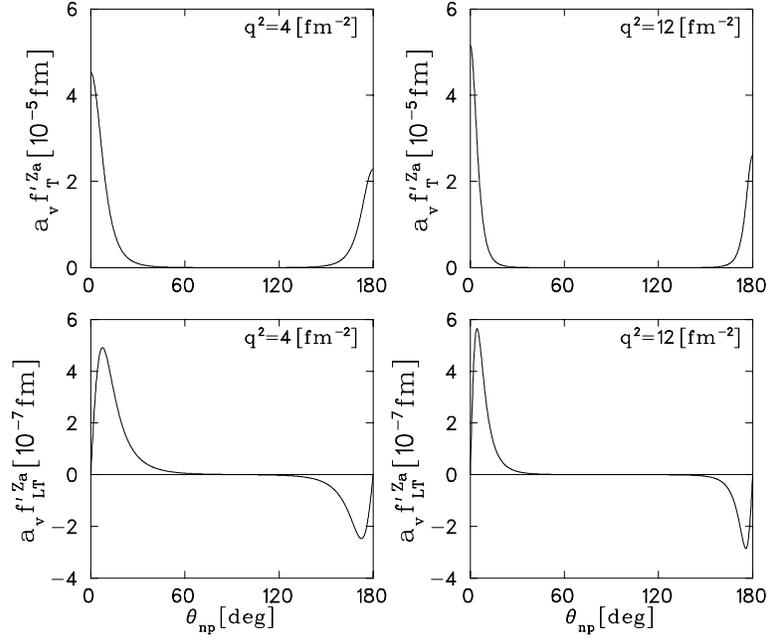,width=10cm,angle=0}}
\vspace*{.5cm}
\caption{
Transverse and longitudinal-transverse pnc structure functions 
from the weak hadronic 
axial current at $\vec q^{\, 2}=4\,\mbox{and}\,12\,\mbox{fm}^{-2}$.
}
\label{fza}
\end{figure}

\begin{figure}
\centerline{\psfig{figure=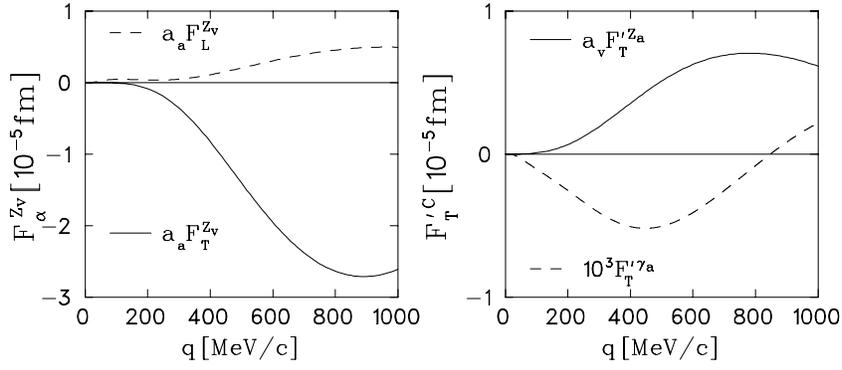,width=11cm,angle=0}}
\vspace*{.5cm}
\caption{
Pnc form factors along the quasifree ridge for the weak hadronic vector 
current (left panel), for the weak hadronic axial current and for the pnc 
deuteron components (right panel).
}
\label{pncff}
\end{figure}

\begin{figure}
\centerline{\psfig{figure=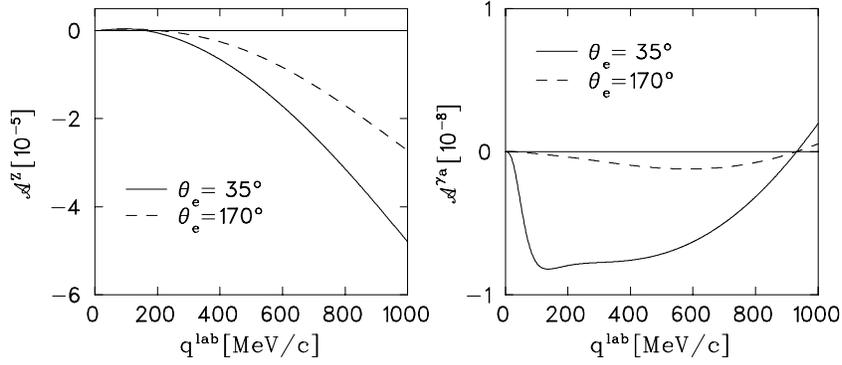,width=11cm,angle=0}}
\vspace*{.5cm}
\caption{
Contributions to longitudinal asymmetry for polarized electrons along the 
quasifree ridge from electroweak interference (left panel) and from the pnc 
deuteron components (right panel) for forward and backward electron 
scattering ($\theta_e=35^\circ\,\mbox{and}\, 170^\circ$) in the laboratory 
frame.
}
\label{longasy}
\end{figure}

\begin{figure}
\centerline{\psfig{figure=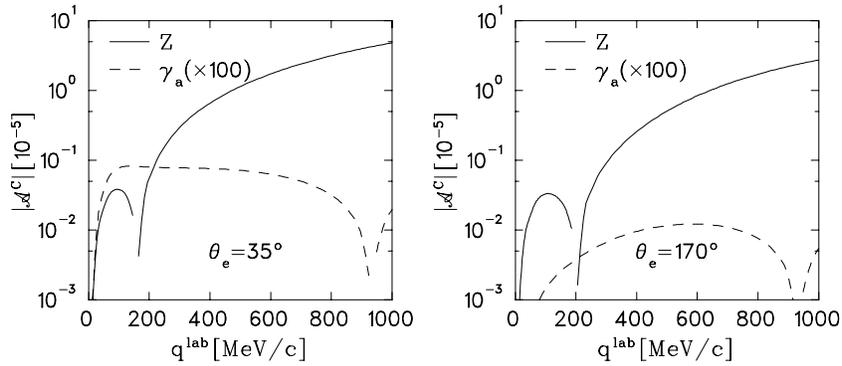,width=11cm,angle=0}}
\vspace*{.5cm}
\caption{
As Fig.\ \protect\ref{longasy} but on a logarithmic scale. 
}
\label{longasylog}
\end{figure}

Note, that the components of the spin operators of both particles refer 
to the frame associated with the final $np$ c.m.\ system denoted 
by $(x,\,y,\,z)$. Its $z$-axis is parallel to
$\vec p_{np}$ in the reaction plane and its $y$-axis parallel
to $\vec q\times \vec p_{np}$ perpendicular to the reaction plane. Thus
the polarization components of particle ``1'' (here the proton)
are chosen according to the Madison convention while for particle ``2''
(neutron) the $y$- and $z$-components
of $\vec P$ have to be reversed in order to comply with this convention.
The spherical angles of proton and neutron momenta
with respect to the reference
frame associated with the reaction plane in the c.m.\ system are
$\theta^{c.m.}_p=\theta$, $\phi^{c.m.}_p=\phi$ and
$\theta^{c.m.}_n=\pi-\theta$, $\phi^{c.m.}_n=\phi+\pi$
(see Fig.\ \ref{kine}).


\begin{table}
\caption{
``Best'' values of the Glashow-Weinberg-Salam model parameters for the 
parity violating $NN$ interaction $V^{pnc}$ from \protect\cite{DDH80} in 
units of $g_\pi=3.8\times 10^{-8}$.
}
\begin{tabular}{cccccc}
$f_\pi$ & $h_\rho^0$ & $h_\rho^1$ & $h_\rho^2$ & 
$h_\omega ^0$ & $h_\omega ^1$ \\
$12$ & $-30$ & $-0.5$ & $-25$ & $-5$ & $-3$\\
\end{tabular}
\label{tab1}
\end{table}

\begin{table}
\caption{Listing of polarization observables $X$ (differential cross 
section $X=1$, and polarization components of outgoing nucleons), separated 
into scalars (set $A$) and pseudoscalars (set $B$).}
\begin{tabular}{ccccccccc}
$A$ & 1 & $y(1)$ & $y(2)$ & $x(1)x(2)$ & $x(1)z(2)$ & $y(1)y(2)$ & 
$z(1)x(2)$ & $z(1)z(2)$\\
$B$ & $x(1)$ & $z(1)$ & $x(2)$ & $z(2)$ & $x(1)y(2)$ & $y(1)x(2)$ & 
$y(1)z(2)$ & $z(1)y(2)$\\
\end{tabular}
\label{tab2}
\end{table}

\end{document}